\documentclass[a4paper,12pt]{article}

\usepackage[T1]{fontenc}
\usepackage{amsmath,amssymb,latexsym}
\usepackage[english]{babel}
\usepackage{graphicx}
\usepackage[T2A]{fontenc}
\usepackage[utf8]{inputenc}
\usepackage{overpic}

\title{Gravity with Higher Derivatives in D-Dimensions}

\author{{Sergey G. Rubin}$^{1,2}$\thanks{e-mail: sergeirubin@list.ru}  \and {Arkadiy Popov}$^{2,*}$\thanks{e-mail: arkady\_popov@mail.ru}  \and {Polina M. Petriakova}$^{1}$\thanks{e-mail: petriakovapolina@gmail.com} }

\date{\em \small
$^{1}$  Moscow Engineering Physics Institute, National Research Nuclear University MEPhI, Kashirskoe shosse 31, 115409 Moscow, Russia;   \\
$^{2}$ N.I. Lobachevsky Institute of Mathematics and Mechanics, Kazan  Federal  University,  \mbox{Kremlevskaya  Street  18},  420008 Kazan,  Russia}

\begin{document}

\maketitle

\begin{abstract}
The aim of this review is to discuss the ways to obtain results based on gravity with higher derivatives in D-dimensional world. We considered the following ways: (1) reduction to scalar tensor gravity, (2) direct solution of the equations of motion, (3) derivation of approximate equations in the presence of a small parameter in the system, and (4) the method of test functions. Some applications are presented to illustrate each method. The unification of two necessary elements of a future theory is also kept in mind---the extra dimensions and the extended form of the gravity.
\end{abstract}


\section{Introduction}

Higher derivative theories of gravity is widely used in modern research despite the internal problems inherent in this approach~\cite{1988PhLB..214..515B, 2015arXiv150602210W}. One such problem is the Ostrogradsky instabilities~\cite{2017PhRvD..96d4035P}.  The~$f(R)$-gravity is the simplest extension of the Einstein--Hilbert theory of gravity, which is free from Ostrogradsky instabilities.
Reviews of $f(R)$---theories, including extension to the Gauss--Bonnet gravity, can be found in~\cite{DeFelice:2010aj, 2011PhR...509..167C}. The~specific form of the function $f(R)$ was considered in~\cite{2003Ap&SS.283..679G,2006GrCo...12..253S,2006A&AT...25..447S}.

Another widespread idea, the~world with extra dimensions is considered as the necessary element for a complete fundamental theory. The~idea of extra dimensions is also used to  explain cosmological evolution~\cite{Abbott:1984ba}.  The~invisibility of extra dimensions can be explained by their small size smaller than $10^{-18}$ cm.

One of the aims of our research is to combine two main elements of future theory---the gravity with higher derivatives and the extra dimensions. The~latter could reveal itself at the inflationary energies and~higher.

It is generally assumed that our Universe was born at Planck energies and evolved by expanding and cooling to its present state. The~inflationary stage is characterized by sub-Planckian energy density and looks inevitable. A~description of the spontaneous creation of the Universe with the inflationary regime can be found in~\cite{Firouzjahi:2004mx}.
Models describing the inflation have been elaborated using many different ingredients, like supersymmetry~\cite{Antusch:2011wu} and attracting the inflationary idea for explanation of another cosmological problems like primordial black holes~\cite{2005APh....23..265K, 2009NuPhB.807..229D}, dark matter~\cite{Gani:2014lka}, and~baryogenesis~\cite{1997PhRvD..56.6155D}. For~review, see, for~example,~\cite{Cline:2018fuq,Akrami:2018odb} and references therein.
Since the energy scale at the inflationary stage of the evolution of the Universe is high enough, some quantum effects may be manifested and responsible for the inflationary regime. Two typical elements of quantum gravity can play a role in the construction of inflationary models: nonlinear geometrical extensions of Einstein's theory of gravity and extra dimensions. Moreover, the~quantization of gravity requires a nonlinear geometric extension of the Einstein--Hilbert action. The~first formulation of the inflationary model, the~Starobinsky model~\cite{Starobinsky:1980te}, considers nonlinear geometric terms belonging to the class of $f(R)$ theories.

To our knowledge, there are four ways to move forward in the framework of higher derivative theories of gravity in D-dimensions:

\begin{enumerate}
\item{Reduction of action to the scalar tensor gravity. This way is effective for the action in the form}
\begin{eqnarray}\label{act1}
&& S=\frac{m_D ^{D-2}}{2}\int d^{D}Z \sqrt{|g_D|}f(R).
\end{eqnarray}

\item{Direct solution to equations of motion.}

\item{Derivation of approximate equations provided that a system contains a small parameter.}

\item{Method of trial functions.}
\end{enumerate}

The next sections are devoted to their~discussion.

Throughout this paper, we use the following conventions:  $R_{ABC}^D=\partial_C\Gamma_{AB}^D-\partial_B\Gamma_{AC}^D+\Gamma_{EC}^D\Gamma_{BA}^E-\Gamma_{EB}^D\Gamma_{AC}^E$,
\ $R_{MN}=R^F_{MFN}$.

\section{Reduction of Action to the Scalar-Tensor~Gravity }

$f(R)$ theories of gravity or, more generally, higher derivative theories of gravity are now widely used as a tool for theoretical research. The~interest in $f(R)$ theories is motivated by inflationary scenarios of the evolution of the Universe, starting with Starobinsky's
pioneering work~\cite{Starobinsky:1980te}.
A number of viable $f(R)$ models in four-dimensional space-time satisfying the observable constraints are discussed in~\cite{2014JCAP...01..008B,2007PhLB..651..224N,Sokolowski:2007rd}.

The first method is based mostly on the conformal transformations that lead to the standard form of Einstein's gravitational action. The~price is the appearance of the additional dynamic variable in the form of the scalar field. Necessary formulas for the D-dim space are represented in the section~below.

\subsection{Conformal Transformations in D~Dimensions}
Action \eqref{act1} can be reduced to a scalar-tensor model in two steps.
Firstly, consider the action depending on auxiliary scalar field $\chi$
\begin{equation}\label{ST}
    S_{ST} = \frac{m_D ^{D-2}}{2} \int  d^{D}Z \sqrt{|g_D|}[f'(\chi)R+f(\chi)-f'(\chi)\chi ]
\end{equation}

One of the classical equation is
\begin{equation}
f''(\chi)(R-\chi)=0
\end{equation}

Thus, $\chi=R$ provided that $f''(\chi)\neq 0$. Substituting this into \eqref{ST}, we arrive to initial action \eqref{act1}. These~actions are equivalent at the classical~level.

As the second step is based on the well known conformal transformation, see e.g.,~\cite{Bronnikov:2003rf}
\begin{equation}\label{gconf}
{g}_{AB}=|\Omega|^{\frac{-2}{D-2}}\hat{g}_{AB},
\end{equation}
which leads to the Ricci scalar transformation in the form
\begin{equation}\label{Rconf}
    \sqrt{g_D}\,\Omega\, R=(sign\, \Omega) \sqrt{\hat{g}_D}
    \left[\hat{R}+\frac{D-1}{D-2}\frac{\partial_A \Omega \hat{g}^{AB}\partial_B \Omega}{\Omega^2}\right] +div.
\end{equation}

Here, all letters with hat are functions of $\hat{g}_{AB}$, and ${div}$ denotes a full divergence which does not contribute to the field~equations.

Application of the conformal transformation \eqref{gconf} with $\Omega=f'$ to expression \eqref{ST}
gives
\begin{equation}\label{ST1}
S=\frac{1}{2}\int d^D x\sqrt{|\hat{g}_D|}[ \hat{R}+(\partial \psi)^2-2V(\psi)]
\end{equation}
where
\begin{equation}
\psi=\sqrt{\frac{D-1}{D-2}}\ln{f'(\chi)}
\end{equation}
($m_D=1$)
and
\begin{equation} \label{V}
V(\psi)=  e^{\frac{-D}{\sqrt{(D-1)(D-2)}}\psi}U(\chi),\quad U(\chi)\equiv \frac12 (f'(\chi)\chi-f(\chi)).
\end{equation}

The Lagrangian containing the Ricci scalar in the form of specific function $f(R)$ is transformed into the scalar-tensor model that strongly simplifies the subsequent calculations. The~applications are wide and can be found in a set of publications~\cite{1999FCPh...20..121F,2007PhLB..651..224N,DeFelice:2010aj,2014JCAP...01..008B,Sokolowski:2007rd}. Here, we consider one of the applications concerning the Starobinsky model of inflation~\cite{Starobinsky:1980te}.

\subsection{The Starobinsky~Model}

This model provides the best fit to the observational data at present times. Historically, this was the first model of inflation, but~many models containing the scalar field(s) were intensively discussed during a couple of decades. These models are well elaborated and intuitively~clear.

The model in question is described by the action \eqref{act1} with $D=4$ and $f(R)=R+R^2/6M^2$. As~one can see from the previous section, the~conformal transformation leads to the scalar-tensor model of inflation, allowing the application of already known results~\cite{2004JHEP...11..063B}. Indeed, action \eqref{ST} is the starting point for the inflationary models based on a scalar field dynamic. According to \eqref{V}, the~potential has the form
\begin{equation} \label{Vstar}
    V(\psi)=\frac34 m_4^2 M^2
\left(1-e^{-\sqrt{2/3}\psi/m_4 }\right)^2
\end{equation}

We have obtained an effective model containing the Einstein--Hilbert action and the scalar field in the standard form. The~scalar field potential possesses a profound minimum at $\psi =0$. The~later is a necessary element responsible for the field oscillations and hence the reheating just after the end of inflation. The~ground state energy should also be zero with great accuracy due to the extreme smallness of the cosmological~constant.

A variety of inflationary models that differ in the form of the potential are described by action~\eqref{ST1}. Most of them are based on the slow motion of fields and should satisfy some conditions which characterized the inflationary stage.
The slow roll parameters are defined as
\begin{eqnarray}
\label{sr1}
\epsilon &=& \frac{1}{2}\biggr(\frac{V'}{V}\biggl)^2, \\
\label{sr2}
\eta &=& \frac{V''}{V},\\
\label{sr3}
\xi^2 &=& \frac{V' V^{'''}}{V^2}.
\end{eqnarray}

The main parameters are the scalar/tensorial relative amplitude $r$ and the scalar spectral index $n_s$, besides~the running of the spectral index ${d n_s}/{d\ln k}$, which are given, in~terms of the slow roll parameters, as
\begin{eqnarray}\label{nsr}
n_s &=& 1 - 6\epsilon + 2\eta,  \\
r &=& 16\epsilon, \label{nsr2} \\
\frac{d n_s}{d\ln k} &=& - 16\epsilon\eta + 24\epsilon^2 + 2\xi^2. \label{nsr3}
\end{eqnarray}
$k$ is the~wavenumber.

According to~\cite{Cline:2018fuq}, the~observed values are as follows:
\begin{equation}\label{nsrObserv}
n_s\simeq0.968,\quad r<0.1,\quad \frac{d n_s}{d\ln k}= -0.0045\pm 0.0067.
\end{equation}

The potential \eqref{Vstar} satisfies all of them for $M\simeq 10^{-5}$. Notice that this potential has only one parameter, $M$, and~nevertheless gives the best fit to the~observations.

We see that the method of conformal transformations allows us to apply the well-known results elaborated for the scalar fields, thus making the physical picture clearer. The~known flaws of this approach are (a) quantum correspondence before and after the conformal transformations is suspicious, (b) treating more general forms of Lagrangian is not~easy.

Let us discuss the second way---numerical solution to the proper system of~equations.

\section{Direct Solution to Equations of~Motion}

We will illustrate this way for two~cases:
\begin{itemize}
  \item $f(R)$ gravity,
  \item $f(R)$ + Gauss--Bonnet gravity.
\end{itemize}

\subsection{$f(R)$ Gravity} \label{21}

Let a $D=1+d_1+d_2$-dimensional space-time $T\times M_{d_1}\times M_{d_2}$ appeared as a result of some quantum processes at high energies. The~probability of these processes is a subtle point, and we do not discuss it in this article. It is supposed that manifolds are born with a random shape. The~entropy growth in $T\times M_{d_1}\times M_{d_2}$ manifold leads to the entropy minimization of subspace $M_{d_2}$ \cite{Kirillov:2012gy}. We begin our research after completing symmetrization and investigate the classical evolution of subspaces $M_{d_1}$ and $M_{d_2}$ whose metric
\begin{equation}\label{metric1}
	ds^2 = dt^2 - e^{2\beta_1(t)}d\Omega_1^2 - e^{2 \beta_2(t)}d\Omega_2^2
	\end{equation}
is assumed to be maximally symmetrical with a positive curvature. In~this section, we consider the action in the form \eqref{act1}.
The equations of this theory are known,
\begin{equation} \label{AB}
	-\frac{1}{2} f(R)\delta_A^B + \Big(R_A^B +\nabla_{A}\nabla^{B} - \delta_A^B\square \Big) f_R = 0.
	\end{equation}
	
The nontrivial equations of system \eqref{AB} are
\begin{eqnarray}\label{11a}
	&& - \frac12 f(R) + f_R \Big[e^{-2\beta_1(t)}(d_1-1)+\ddot{\beta}_1+\dot{\beta}_1(d_1\dot{\beta}_1 +d_2\dot{\beta}_2) \Big] \nonumber \\
    && + \Big[(1-d_1 )\dot{\beta}_1 - d_2 \dot{\beta}_2   \Big] f_{RR}\dot{R} - f_{RRR} \dot{R}^2 - f_{RR} \ddot{R} = 0,
	\end{eqnarray}
\begin{eqnarray}\label{22a}
	&& -\frac12 f(R)+f_R \Big[e^{-2\beta_2(t)}(d_2-1)+\ddot{\beta}_2+\dot{\beta}_2(d_1\dot{\beta}_1 +d_2\dot{\beta}_2) \Big] \nonumber \\
    && +\Big[(1-d_2 )\dot{\beta}_2 - d_1 \dot{\beta}_1  \Big]f_{RR}\dot{R}-f_{RRR}\dot{R}^2 - f_{RR}\ddot{R} =0,
	\end{eqnarray}
\begin{equation}\label{00a}
	-\frac12 f(R)+\left[d_1 \ddot{\beta}_1 +d_2 \ddot{\beta}_2 +d_1{\dot{\beta}_1}^2 + d_2 {\dot{\beta}_2}^2  \right]f_R
	-\left( d_1 \dot{\beta}_1 +d_2 \dot{\beta}_2 \right)f_{RR}\dot{R} =0
	\end{equation}
in  terms  of  metric \eqref{metric1}.
 Here, we took into account that $\partial_t f_R=f_{RR}\dot{R}$ and $\partial^2_t f_R=f_{RRR}\dot{R}^2 + f_{RR}\ddot{R}$.
The Ricci scalar is
\begin{eqnarray} \label{Ra}
	&& R = d_1\ddot{\beta}_1 + d_2\ddot{\beta}_2
	 + d_1\dot{\beta}_1^2 +  d_2\dot{\beta}_2^2 +  d_1 \left[ e^{-2\beta_1(t)}(d_1 -1)+\ddot{\beta}_1
    \right. \nonumber \\ && \left. \frac{}{}
    +\dot{\beta}_1 (d_1\dot{\beta}_1 + d_2\dot{\beta}_2)\right] + d_2 \Big[ e^{-2\beta_2(t)}(d_2 -1) +\ddot{\beta}_2 + \dot{\beta}_2(d_1\dot{\beta}_1+d_2\dot{\beta}_2) \Big].
	\end{eqnarray}
	
The system of Equations~(\ref{11a})--(\ref{00a}) is a system of fourth-order differential equations with respect to unknown functions $\beta_1(t)$ and $\beta_2(t)$. Note that this system is not solvable with respect to the highest derivatives $\ddddot{\beta_1}$ and $\ddddot{\beta_2}$ because these derivatives appear when expression \eqref{Ra} is substituted into the similar terms $- f_{RR} \ddot{R}$ of Equations \eqref{11a} and \eqref{22a}. Therefore, two equations contain the forth-order derivatives with
equal~multipliers.


Another way to solve the problem is as follows:
we can consider $R(t)$ as an additional unknown function, and~the expression \eqref{Ra} as the fourth equation of the system. In~this case, we obtain a system of second order ordinary differential equations for the unknown functions $\beta_1(t), \beta_2(t)$, and~$R(t)$. Three~equations of this system (for example, \eqref{11a}, \eqref{22a}, \eqref{Ra})
can be solved with respect to $\ddot{\beta}_1, \ddot{\beta}_2, \ddot{R}$.
Then, substitution $\ddot{\beta}_2$ and $\ddot{\beta}_1$ into Equation \eqref{00a} gives equation
\begin{eqnarray} \label{eq00a}
&& - 10 \left[ d_1 \dot{\beta}_1 + d_2 \dot{\beta}_2 \right]f_{RR}\dot{R} +\left[5 R -5 d_1(d_1 -1){\dot{\beta}_1}^2  -10 d_1 d_2 \dot{\beta}_1 \dot{\beta}_2
-5 d_2(d_2 -1){\dot{\beta}_2}^2
\right. \nonumber \\ && \left.
 -5 d_1(d_1 -1) e^{-2\beta_1} -5 d_2(d_2 -1) e^{-2\beta_2}\right]f_R
 -5 f(R) =0,
\end{eqnarray}
which plays the role of restriction for our system of differential equations.
Equation \eqref{eq00a} should be used to obtain a  relation between the initial data $\beta_1 (t_0), \dot{\beta}_1 (t_0), \beta_2 (t_0), \dot{\beta}_2 (t_0)$, and~$\dot{R}(t_0)$. The~value $R(0)$ can be found from auxiliary condition \eqref{eq00a} at $t=0$. Let
\begin{equation}\label{f}
f(R)=a_3R^3 + a_2R^2 + a_1R+a_0
\end{equation}
with the parameter values (the choice is quite arbitrary)
\begin{equation}
a_0=0.6, a_1=1, a_2=-2, a_3=1.3.
\end{equation}

\begin{figure}[h!]
\centering
\includegraphics[width=4.8cm]{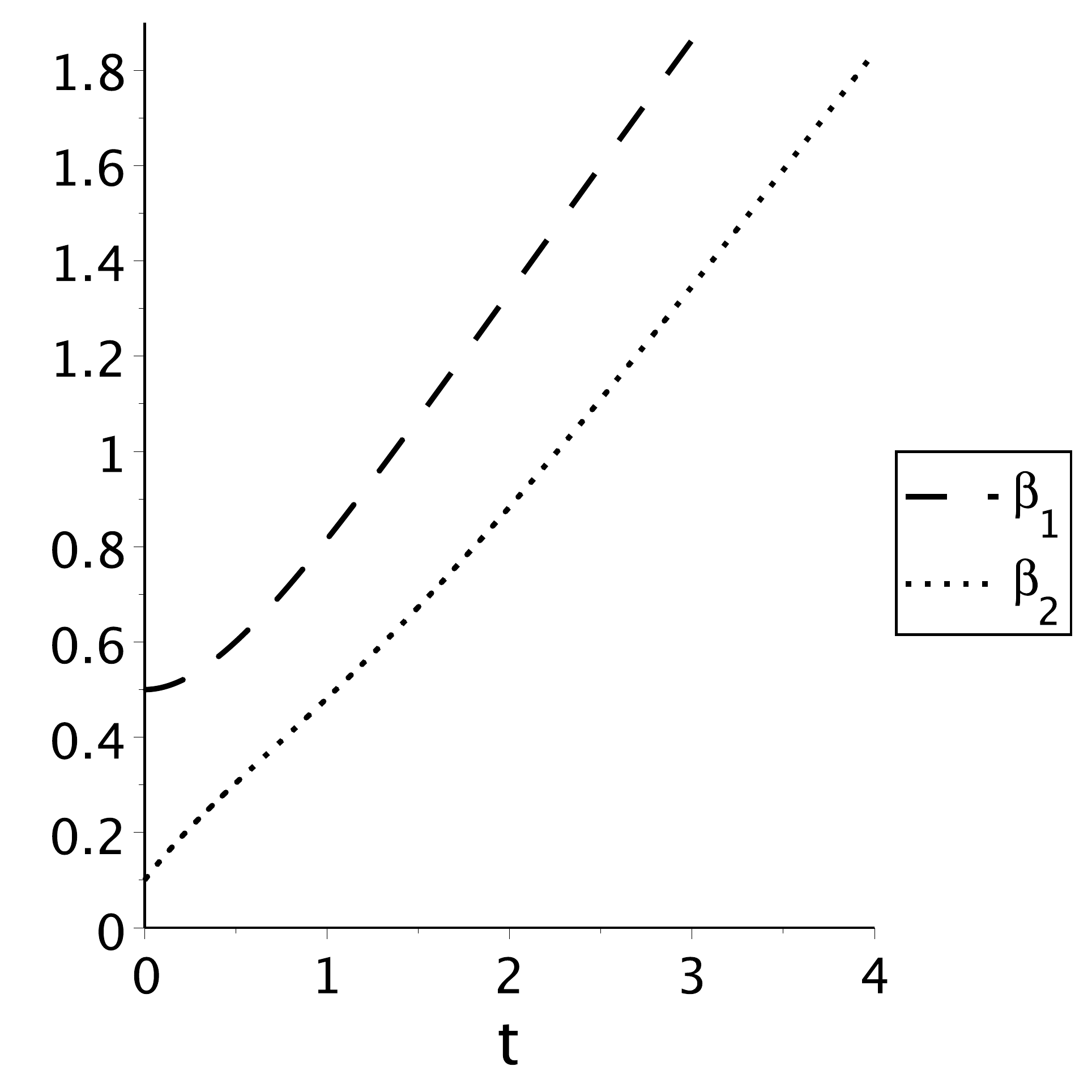}  \quad \includegraphics[width=4.8cm]{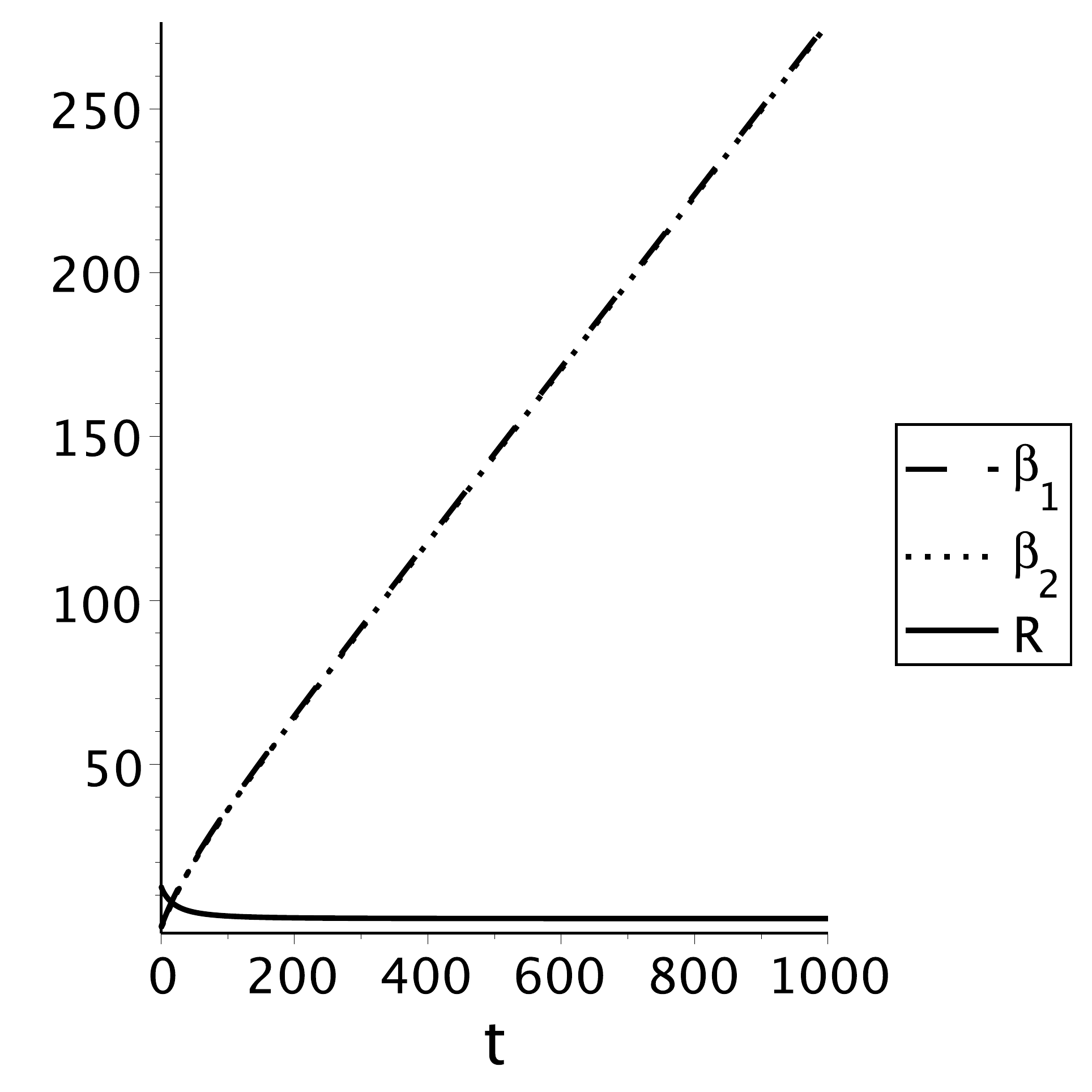}   \quad \includegraphics[width=4.8cm]{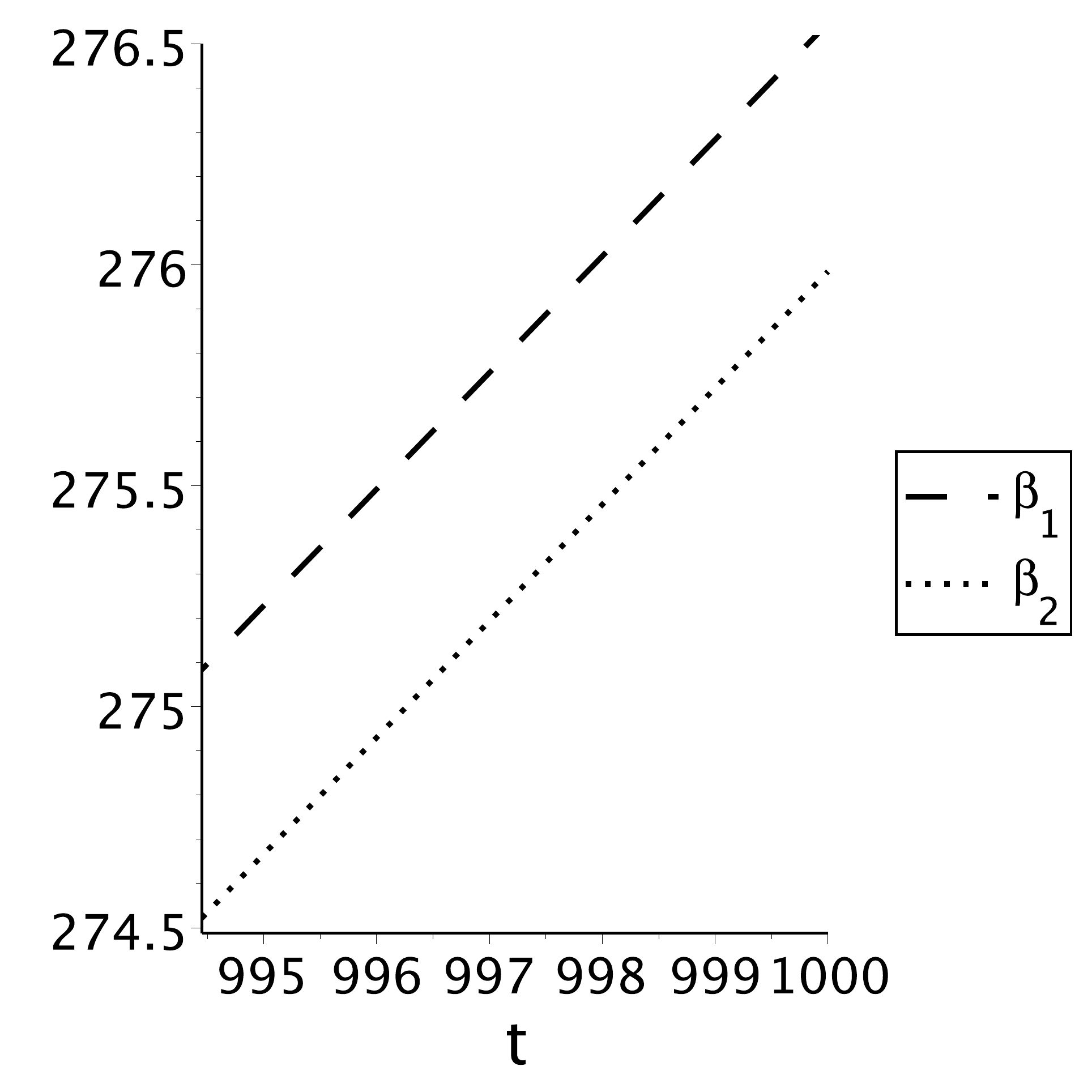}
\caption{Numerical solution of \eqref{11a}, \eqref{22a} and \eqref{Ra} for  $ d_1=d_2=3, \beta_1(0)=0.5, \ \dot{\beta}_1(0)=0, \  \beta_2(0)=0.1, \ \dot{\beta}_2(0)=0.5, \ \dot{R}(0)=0$. $R(0) \simeq 12.67452$ is found from Equation \eqref{eq00a}. }
\label{fig2}
\end{figure}

Figure~\ref{fig2} shows the numerical solution of the system of Equations \eqref{11a}, \eqref{22a}, and \eqref{Ra} in different time ranges. The~region of small values of time is shown at the left panel. The~time behavior of both functions $\beta_1(t)$ and $\beta_2(t)$ differ from each other due to differences in the initial conditions, see the capture of Figure~\ref{fig2}. The~panel in the middle indicates similar asymptotic behavior of the solutions. The~more detailed figure on the right panel helps distinguish the functions $\beta_1(t)$ and $\beta_2(t)$.

The asymptotics of the solution
\begin{equation} \label{asHH}
	\beta_1(t) = H_1 t,\quad 	\beta_2(t) = H_2 t \quad (H_1>0, \ H_2 >0).
\end{equation}
are determined by the Equations
\eqref{11a}--\eqref{Ra} at $t \rightarrow \infty$
\begin{equation} \label{Ras1}
R(t)=d_1 (d_1+1)H_1^2 + d_2 (d_2+1)H_2^2 + 2d_1d_2 H_1H_2 \equiv R_0,
\end{equation}
\begin{eqnarray}\label{space}
\left. f_R \Big(d_1 H_1^2 + d_ 2H_1 H_2  \Big) -\frac12 f \ \right|_{R=R_0}&=&0,\nonumber \\
\left. f_R \Big(d_2 H_2^2 + d_1 H_1 H_2  \Big) -\frac12 f \ \right|_{R=R_0}&=&0, \\
\left. f_R \Big( d_1 {H_1}^2 +d_2 {H_2}^2 \Big) -\frac12 f \ \right|_{R=R_0}&=&0.
\end{eqnarray}

According to these equations, the~subspaces are expanded with equal speed,
\begin{eqnarray}\label{ss1}
H_1=H_2= \left. \sqrt{ \frac{f}{2 (d_1 +d_2) f_R }}\ \right|_{R=R_0}.
\end{eqnarray}

Using this expression, we obtain the equation for $ R_0 $ from \eqref{Ras1}
\begin{eqnarray}
\left.  - (d_1 +d_2 +1) f\frac{}{}\right|_{R=R_0}
+\left. 2  R f_R\frac{}{}\right|_{R=R_0} = 0.
\end{eqnarray}

\begin{figure}[h!]
\begin{center}
\includegraphics[width=6cm]{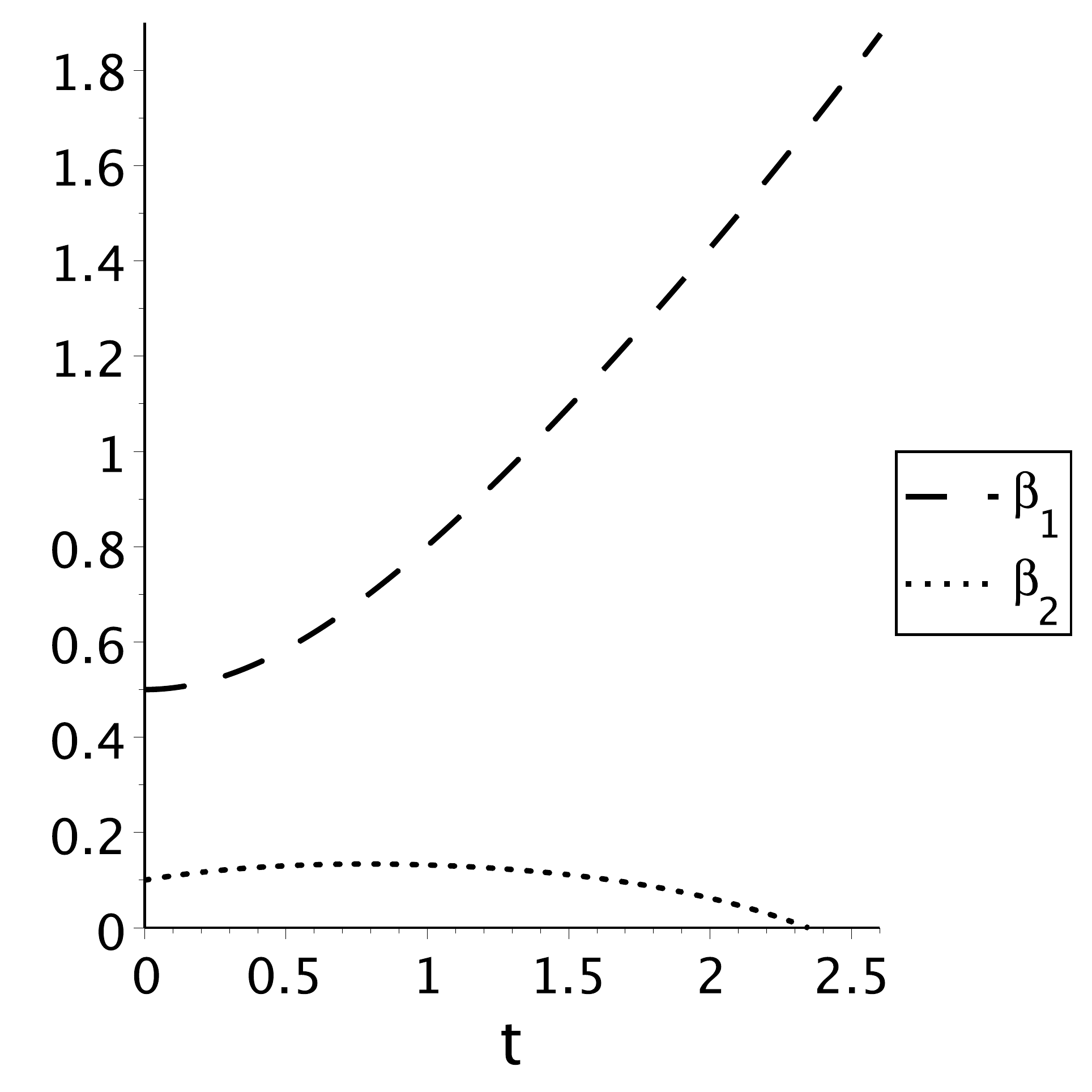}
\caption{The same as in Figure~\ref{fig2} except $\dot{\beta}_2(0)=0.1$.}
\label{fig2a}
\end{center}
\end{figure}

We note that the results at $t \rightarrow \infty$ do not depend on the initial conditions and $H_1, H_2, R_0$ are consistent with the numerical results. Nevertheless, the~conclusion that the asymptotic behavior is independent of the initial conditions in their whole range is hasty.  As~shown in Figure~\ref{fig2a}, for some initial conditions, there is no stable~solution.

The discussion above indicates that time dependence of an extra space metric is determined by the initial conditions. The~latter leads to growing volumes of both extra spaces, which means that such solutions are hardly applicable to the description of our~Universe.

In a more realistic situation, the~radius of one of the subspaces (say, $ M_ {d_2} $) remains constant
\begin{equation}\label{bc}
    \beta_2(t)=\beta_c =const,
	\end{equation}
while the other one, $M_{d_1}$, expands.
In this case, the~system of Equations \eqref{11a}--\eqref{Ra} admits an analytic solution. More definitely, the~combination
$d_1 \cdot$ \eqref{11a} $ -\,d_1 \cdot$\eqref{22a}  + \eqref{00a} $ - f_R \cdot $\eqref{Ra}
gives
\begin{equation}\label{com}
	-\frac{f(R)}{2} +\left[R(t) +\frac{d_1+d_2-d_2(d_1+d_2)}{e^{\beta_c}} \right]f_R=0.
	\end{equation}
	
It means that
\begin{equation} 
R (t) = R_0 = \mbox{const} \,
\end{equation}
and $R_0$ can be found from Equation \eqref{com}.
Then, Equations \eqref{11a}--\eqref{Ra} can be rewritten as
\begin{eqnarray} \label{Rc}
	&&R_0 = 2d_1\ddot{\beta}_1 +d_1(d_1+1) \, \dot{\beta}_1^2 +d_1 (d_1 -1)  \, e^{-2\beta_1(t)} +d_2 (d_2 -1) \, e^{-2\beta_c},
	\end{eqnarray}
\begin{eqnarray}\label{11c}
	&& -\frac12 f(R_0)+\left. f_R \left[e^{-2\beta_1(t)}(d_1-1)+\ddot{\beta}_1+d_1\dot{\beta}_1^2 \right]\right|_{R=R_0} =0,
	\end{eqnarray}
\begin{eqnarray}\label{22c}
	&& -\frac12 f(R_0) +\left. e^{-2\beta_c}(d_2-1) f_R\right|_{R=R_0} =0,
	\end{eqnarray}
\begin{equation}\label{00c}
	-\frac12 f(R_0) +\left. d_1\left( \ddot{\beta}_1 +{\dot{\beta}_1}^2 \right) f_R \right|_{R=R_0}=0.
	\end{equation}
	
Subtracting Equation \eqref{11c} from \eqref{00c}, we obtain
\begin{eqnarray}\label{comc}
	(d_1-1)\left( \ddot{\beta}_1 - \left. e^{-2\beta_1(t)}\right) f_R \right|_{R=R_0} & = & 0,
	\end{eqnarray}
which gives the connection
\begin{eqnarray} \label{ddb1}
    \ddot{\beta}_1 &=& \, e^{-2\beta_1(t)}\,
	\end{eqnarray}
for
$ d_1 \neq 1, \ \left. f_R \right|_{R=R_0} \neq 0 $.
Then, Equations \eqref{Rc}--\eqref{00c} are reduced to
\begin{eqnarray}\label{eq21}
e^{-2\beta_1(t)} + \dot{\beta}_1^2  &=& \frac{e^{-2\beta_c} (d_2-1) }{d_1}=\left.  \frac{ f(R)} {2 d_1 f_R} \right|_{R=R_0}
    =\frac{R_0 -d_2(d_2-1) e^{-2\beta_c}}{d_1(d_1+1)} \equiv H^2.
	\end{eqnarray}
	
The solution of Equations \eqref{eq21} and \eqref{ddb1} with respect to $\beta_1(t)$ is
\begin{eqnarray}\label{b1bc}
    {\beta}_1(t) & = & \pm H\left( t -t_0 \right) +\ln \left( \frac{1 +e^{\mp 2 H (t-t_0)}}{2 H}  \right), \quad H>0,
	\end{eqnarray}
where $H, \ R_0$ and $\beta_c$ can be found from the last relations \eqref{eq21}.

In this section, we have obtained a set of numerical and analytical solutions in $f (R)$ gravity. Dependence on the initial conditions appears to be nontrivial.
The relation between the initial conditions and the asymptotes of the solutions is also not~clear.

\subsection{Starobinsky Model, Direct~Calculation}

Let us come back to Starobinsky model discussed above and perform the direct simulations.
Both~approaches should give similar results. To~this end, consider the four-dimensional theory
described by the action
\begin{equation}\label{act0}
S[g_{\mu \nu}]=\frac12 m_{Pl}^{2}\int d^4 x \sqrt{|g_4|}\, f(R) \, .
\end{equation}

The corresponding equations of motion in four dimensions are as follows:
\begin{equation}\label{eqf(R)}
f ' _{R} (R) R_{\mu \nu} - \frac{1}{2} \, f(R) g_{\mu \nu} + \Bigl[ \nabla_{\mu} \nabla_{\nu} - g_{\mu \nu} \Box \Bigr]f ' _{R} (R) =0 \, , \quad \Box \equiv g^{\mu \nu} \nabla_{\mu} \nabla_{\nu} \,
\end{equation}
with $\mu , \nu =1,2,3,4$.
Taking into account the choice of the metric of the three-dimensional sphere
\begin{equation}\label{ds4}
    ds^2=g_{\mu\nu}dx^\mu dx^\nu= dt^2 - \text{e}^{2 \alpha(t)}\Bigl(dx^2 + \sin^2{x} \,  dy^2 + \sin^2{x} \, \sin^2{y} \,  dz^2\Bigr)
\end{equation}
we obtain the system of equations
\begin{eqnarray}\label{eqsyst1}
\begin{cases}
6\dot{\alpha}\dot{R} f''_R (R)-6 \Bigl(\ddot{\alpha} +\dot{\alpha}^2  \Bigr)f'_R (R) +f(R) = 0 \, ,\\
2\dot{R}^2 f'''_R (R)+2\Bigl(\ddot{R} +2\dot{\alpha} \dot{R} \Bigr)f''_R (R)-\Bigl(2\ddot{\alpha} +6\dot{\alpha}^2 +4\text{e}^{-2\alpha } \Bigr)f'_R (R) + f(R)=0 \, ,
\end{cases}
\end{eqnarray}
where the definition of the Ricci scalar for metric \eqref{ds4} is
\begin{equation} \label{RR}
    R(t)= 12\dot{\alpha}^2  + 6\ddot{\alpha}  + 6\text{e}^{-2\alpha }.
\end{equation}

Let us choose the form of the function as in the Starobinsky model
\begin{equation}
f(R)=\cfrac{R^2}{6M^2}+R \, ,
\end{equation}
where $M$ is a constant parameter with dimension of mass and is defined as
\begin{equation}\label{m}
M \sim 1.5\cdot 10^{-5} \, m_{Pl}\, \biggl(\cfrac{50}{N_e}\biggr), \quad N_e = 55 \div 60 \, .
\end{equation}

Using the second equation of the system \eqref{eqsyst1} and the definition of the Ricci scalar \eqref{RR}, we get
\begin{eqnarray} \label{sS3}
\begin{cases}
\ddot{R} = - 2 \dot{\alpha}(t) \dot{R}   - \cfrac{1}{12} \,R^2  + \Bigl( \dot{\alpha}^2  + \text{e}^{-2\alpha } -M^2\Bigr)R  + 3M^2 \Bigl( \dot{\alpha}^2 + \text{e}^{-2\alpha }\Bigr) , \\
\ddot{\alpha} = - 2\dot{\alpha}^2  - \text{e}^{-2\alpha } + \cfrac{1}{6}\, R  \, .
\end{cases}
\end{eqnarray}

The second derivative $ \ddot{\alpha}(t) $ is disappeared in a combination of the first equation of the system \eqref{eqsyst1} and definition of the Ricci scalar \eqref{RR}. Finally, we obtain the quadratic equation
\begin{equation}\label{R(t)_eqsq}
R^2 -12\bigl(\dot{\alpha}^2 +\text{e}^{-2\alpha }\bigr)R  - 12 \dot{\alpha}  \dot{R}  - 36 M^2 \bigl( \dot{\alpha}^2 +\text{e}^{-2\alpha } \bigr) =0 \, .
\end{equation}

This expression being substituted into the first equation of the system \eqref{sS3}, leads to the equation of free damped harmonic oscillations
\begin{equation}\label{Rdd}
  \ddot{R}  + 3 \dot{\alpha} \dot{R}  + M^2 R   = 0\, .
 \end{equation}

 We need to choose the initial conditions to find a solution for the unknown functions $\alpha(t)$ and $R(t)$. Let the initial conditions on the function $ \alpha(t) $ in Planck units be given as
\begin{equation}\label{StAlph}
 \alpha(0) \equiv \alpha_0 = - \ln{H_{infl}} = \ln{10^6} \sim 13.8 \, , \quad \dot{\alpha}(0) \equiv \alpha_1 = H_{infl} \sim 10^{-6}.
\end{equation}

Solving the previously obtained Equation \eqref{R(t)_eqsq}, we find an expression for the function $ R(t) $ at the initial time depending on the value of other initial conditions
\begin{equation}\label{R(t)_sq0}
R(0) \equiv R_0 = 6\bigl(\alpha_1^{2}+\text{e}^{-2\alpha_0} \bigr) \pm  \sqrt{\,36\Bigl(\alpha_1^{2}+\text{e}^{-2\alpha_0} \Bigr)^2 + \Bigl(12\alpha_1 R_1 +36 M^2\,\bigl(\alpha_1^{2}+\text{e}^{-2\alpha_0} \bigr) \Bigr)} \,\, .
\end{equation}

Thus, after~substitution of specific values of quantities, we get
\begin{equation}\label{StR}
 \dot{R}(0) \equiv R_1 = 0\,  \xrightarrow{\eqref{R(t)_sq0}} \, R_{+}(0) = 2.9 \cdot 10^{-11}, \, R_{-}(0) = -4.3 \cdot 10^{-12}.
\end{equation}

\begin{figure}[h!]
 \begin{center}
  \includegraphics[width=0.4\textwidth]{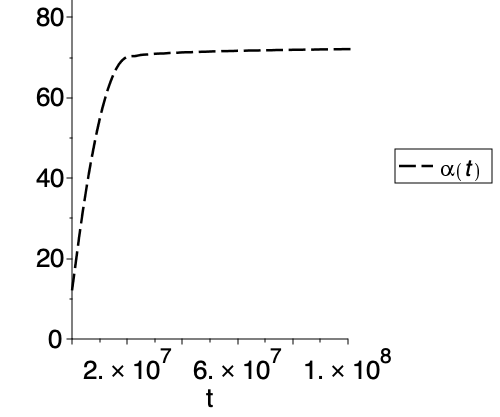} \
  \includegraphics[width=0.4\textwidth, height=4.5cm]{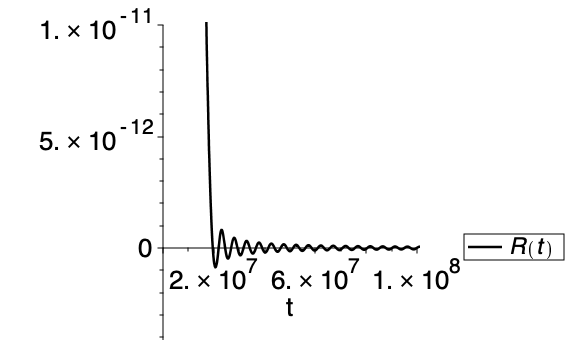}
 \end{center}
 \caption{Solution for the system \eqref{sS3} with the initial conditions \eqref{StAlph} and \eqref{StR}.}
 \label{Fig:Starob}
\end{figure}

The result of numerical solving the system \eqref{sS3} with the choice of the initial conditions \eqref{StAlph} and \eqref{StR} is presented in Figure~\ref{Fig:Starob}. The~time interval of inflation is between $ 10^{-42} \, \text{s} $ and $ 10^{-36} \, \text{s} $ and is equal on the Planck scale to $\sim$20  and $ \sim$2 $\cdot 10^{7} $, respectively. The~slope of the straight line for the function $ \alpha(t) $ changes at a value of time close to the end of inflation, where oscillations of the curvature $ R(t) $ begin. This is shown on the right side of Figure~\ref{Fig:Starob}.
The origin of these oscillations is clear from the Equation \eqref{Rdd}. When a regime $ 1.5 \dot{\alpha}(t) < M $ occurs in this equation, the~solution is damped oscillations. This condition leads to the spacetime being considered flat or of small curvature: at the inflation stage, $ R \sim 12H_{Infl}^{2} = 12 \dot{\alpha}^{2}(t) \Rightarrow |R| << M^2\sim 10^{-10}$. The~size of the space at the inflation stage should increase by $ \sim 10^{26} $ times and, for the function $ \alpha(t) $, it is a value of about $ 60 $; we see that the result shown in Figure~\ref{Fig:Starob} confirms this fact. Numerical calculation gives the value of the Hubble constant in Planck units $ H_{Infl} \equiv \dot{\alpha}(t) \sim 10^{-6} $ within the required order during~inflation.
Asymptotic behavior can be found after reconstruction of the numerical solution. The~size of the sphere $\sim \exp\{\alpha(t_{Univ})\} \sim \exp\{154\} $, the~curvature value $ R(t_{Univ}) \sim 10^{-122} $ and the value of the derivative of the function $\dot{\alpha} (t_{Univ}) \sim 10^{-61} $.

As a result, the~direct numerical simulation and application of the conformal transformation give similar results and may be used if a Lagrangian is not very~complicated.

\subsection{$f(R)$ + Gauss--Bonnet Gravity}\label{eq}

In this section, we extend our study to action in the more complex form
\begin{eqnarray}\label{Seff}
&&S_{eff}=\frac12 m_D^{D-2}\int d^D x \sqrt{|g_D|}\Big[f(R)+ c_1 R_{AB}R^{AB} + c_2 R_{ABCD}R^{ABCD}\Big].
\end{eqnarray}

This action represents an example of the effective theory~\cite{2007ARNPS..57..329B}.
Here, $c_1,c_2$ are parameters of the Lagrangian, and~$f(R)$ is a function of the Ricci scalar $R$. The~Gauss--Bonnet Lagrangian
\begin{eqnarray}
\label{L_GB}
{\cal L}_{GB} = k\sqrt{-g}\biggr\{R^2 - 4R_{AB}R^ {AB} + R_{ABCD}R^{ABCD}\biggl\}
\end{eqnarray}
belongs to such set of models and is the appropriate starting point because of the absence of higher~derivatives.

The action
\begin{equation}\label{Sgen}
S_{gen}=\frac12 m_D^{D-2}\int d^D x \sqrt{|g_D|}\Big[\tilde{f}(R)+k\biggr\{R^2 - 4R_{AB}R^{AB} + R_{ABCD}R^{ABCD}\biggl\}\Big]
\end{equation}
is used in the following. This action is the particular case of the action \eqref{Seff} provided that
 $\tilde{f}(R)=f(R) - kR^2$ and $c_1=-4k, c_2 =k$. In~what follows, we will consider only the quadratic function
\begin{equation}
 f(R)=  aR^2+bR+c
 \end{equation}
 ($b=1$ without the loss of generality).

 We assume that both subspaces are three-dimensional maximally symmetric subspaces of positive~curvature
\begin{eqnarray}\label{metric2}
ds^2 = &&dt^2-e^{2\alpha(t)} m_D^{-2} [dx^2 +\sin^2(x)dy^2 +\sin^2(x)\sin^2(y)dz^2]
\nonumber\\&& \quad
-e^{2\beta(t)} m_D^{-2} [d\theta^2 + \sin^2(\theta)\, d\phi^2+ \sin^2(\theta) \sin^2(\phi)\, d\psi^2].
\end{eqnarray}

Einstein's equations for this model are
\begin{eqnarray}\label{AB_}
	&&-\frac{1}{2}\tilde{f}(R)\delta_B^A + (R_B^A +\nabla^{A}\nabla_{B} - \delta_B^A \square) \tilde{f}_R
    +k\Big[ -8 {R^{AC}}_{;BC}  -12R^{AC} R_{CB}
    \nonumber \\ &&
     +2\delta^A_B R_{CD} R^{CD}
    -\frac{\delta^A_B}{2}  R^{CDEF} R_{CDEF} +2 R^{ACDE} R_{BCDE}  +4 {R^{;A}}_{;B}
    \nonumber \\ &&
     +4 {R^{CAD}}_B^{\ \cdot} \ R_{CD} -\frac{\delta^A_B}{2} R^2 +2R R^A_B \Big] = 0,
	\end{eqnarray}

The nontrivial system of Equation \eqref{AB_} is
\begin{eqnarray}\label{tt2}
	&& -36 k \Big[ {\dot{\alpha} {\dot{\beta}}^3 +3 {\dot{\alpha}}^2 {\dot{\beta}}^2
    + \dot{\alpha}}^3 \dot{\beta}
    +e^{-2 \alpha} \dot{\beta} \left( \dot{\beta}  + \dot{\alpha} \right)
    +e^{-2 \beta} \dot{\alpha} \left( \dot{\beta} + \dot{\alpha} \right)
    + e^{-2 \beta} e^{-2 \alpha}
    \Big]
    \nonumber \\ &&
    -3 \tilde{f}_{RR} \dot{R} \left( \dot{\alpha} +\dot{\beta} \right)
    +3 \tilde{f}_R \left( \ddot{\alpha} + \ddot{\beta} +{\dot{\alpha}}^2 +{\dot{\beta}}^2 \right) -\frac12 \tilde{f}(R)=0,
	\end{eqnarray}
\begin{eqnarray}\label{xx2}
	&& -12 k \left\{ 2 \ddot{\alpha} \dot{\beta}\left( \dot{\alpha} + \dot{\beta} \right)
    +\ddot{\beta} \left( {\dot{\alpha}}^2 +4 \dot{\alpha}\dot{\beta} +{\dot{\beta}}^2 \right)
    +2 {\dot{\alpha}}^3 \dot{\beta} +6 {\dot{\alpha}}^2 {\dot{\beta}}^2 + 6 \dot{\alpha} {\dot{\beta}}^3 +{\dot{\beta}}^4
    \right. \nonumber \\ &&  \left.
    +e^{-2 \alpha} \left[ \ddot{\beta} +2{\dot{\beta}}^2 \right]
    +e^{-2 \beta} \left[ 2 \ddot{\alpha} +\ddot{\beta} + 3{\dot{\alpha}}^2 +2 \dot{\alpha}\dot{\beta} +{\dot{\beta}}^2 \right] +e^{-2 \alpha} e^{-2 \beta}
    \right\}
    - \tilde{f}_{RRR} {\dot{R}}^2
    \nonumber \\ &&
    - \tilde{f}_{RR} \left[\ddot{R} +\dot{R} \left(2 \dot{\alpha} +3\dot{\beta} \right) \right]
    + \tilde{f}_R \left( \ddot{\alpha} +3{ \dot{\alpha}}^2 + 3 { \dot{\alpha}} \dot{\beta} +2 e^{-2 \alpha} \right) -\frac12 \tilde{f}(R)=0,
	\end{eqnarray}
\begin{eqnarray}\label{thth2}
	&& -12 k \left\{ \ddot{\alpha} \left( {\dot{\alpha}}^2  +4\dot{\alpha} \dot{\beta} + {\dot{\beta}}^2 \right) +2 \ddot{\beta} \dot{\alpha} \left( \dot{\alpha}+ \dot{\beta} \right)
    +{\dot{\alpha}}^4 +6 {\dot{\alpha}}^3 \dot{\beta} +6 {\dot{\alpha}}^2 {\dot{\beta}}^2 + 2 \dot{\alpha} {\dot{\beta}}^3
    \right. \nonumber \\ &&  \left.
    +e^{-2 \alpha} \left[ \ddot{\alpha} +2\ddot{\beta}+{\dot{\alpha}}^2 +2 \dot{\alpha}\dot{\beta} +3{\dot{\beta}}^2 \right]
    +e^{-2 \beta} \left[ \ddot{\alpha} +2{\dot{\alpha}}^2 \right] +e^{-2 \alpha} e^{-2 \beta}
    \right\}
    - \tilde{f}_{RRR} {\dot{R}}^2
    \nonumber \\ &&
    - \tilde{f}_{RR} \left[\ddot{R} +\dot{R} \left(3 \dot{\alpha} +2\dot{\beta} \right) \right]
    + \tilde{f}_R \left( \ddot{\beta} + 3 { \dot{\alpha}} \dot{\beta} +3{ \dot{\beta}}^2  +2 e^{-2 \beta} \right) -\frac12 \tilde{f}(R)=0,
	\end{eqnarray}
where we have kept in mind denotations $\partial_t ( \tilde{f}_R )=\tilde{f}_{RR}\dot{R}$ and $\partial^2_t ( \tilde{f}_R) = \tilde{f}_{RRR}\dot{R}^2 + \tilde{f}_{RR}\ddot{R}$.
The Ricci scalar is
\begin{eqnarray}\label{R}
	&& R = 6 \left(\ddot{\beta} +  \ddot{\alpha} + 2 {\dot{\beta}}^2 + 3 \dot{\alpha} \dot{\beta} + 2 {\dot{\alpha}}^2 + e^{-2 \beta} + e^{-2 \alpha} \right).
	\end{eqnarray}
	
Here, and in the following, the units $m_D=1$ are~assumed.
	
As in Section \ref{21}, it is convenient to consider the Ricci scalar $R(t)$ as the additional unknown function and interpret definition \eqref{R} as the fourth equation  of the system with respect to unknown functions $\alpha(t), \beta(t)$, and~$R(t)$.
Three equations of this system (for example, \eqref{xx2}--\eqref{R})
can be solved with respect to the higher derivatives $\ddot{\alpha}, \ddot{\beta}, \ddot{R}$.
Then, substituting $\ddot{\alpha}$ and $\ddot{\beta}$ into Equation \eqref{tt2}, we obtain the equation
\begin{eqnarray} \label{eq002}
&& -36 k \left[ {\dot{\alpha}}^3 \dot{\beta} +3{\dot{\alpha}}^2 {\dot{\beta}}^2 +{\dot{\alpha}} {\dot{\beta}}^3
+e^{-2\alpha}\dot{\beta}\left(\dot{\alpha} +\dot{\beta} \right) +e^{-2\beta}\dot{\alpha}\left(\dot{\alpha} +\dot{\beta} \right)
+e^{-2\alpha} e^{-2\beta}\right]
\nonumber \\ &&
-3\left( \dot{\alpha} +\dot{\beta} \right)\dot{R} \tilde{f}_{RR}
+\left(-3{\dot{\alpha}}^2 -9\dot{\alpha}\dot{\beta} -3\dot{\beta} -3e^{-2\alpha} -3e^{-2\beta} +\frac{R}{2}  \right)\tilde{f}_R -\frac{\tilde{f}}{2}
=0,
\end{eqnarray}

The standard initial conditions are not independent due to Equation \eqref{eq002}. The~latter must be used to obtain a relation between these initial~data.

The free parameters $a\sim k\sim m_D^{-2}, c\sim  m_D^{2} $ are not related to the observational values because of the strong and uncontrolled contribution of the quantum corrections at sub-Planckian~energies.

\begin{figure}[h!]
\centering
\includegraphics[width=7cm]{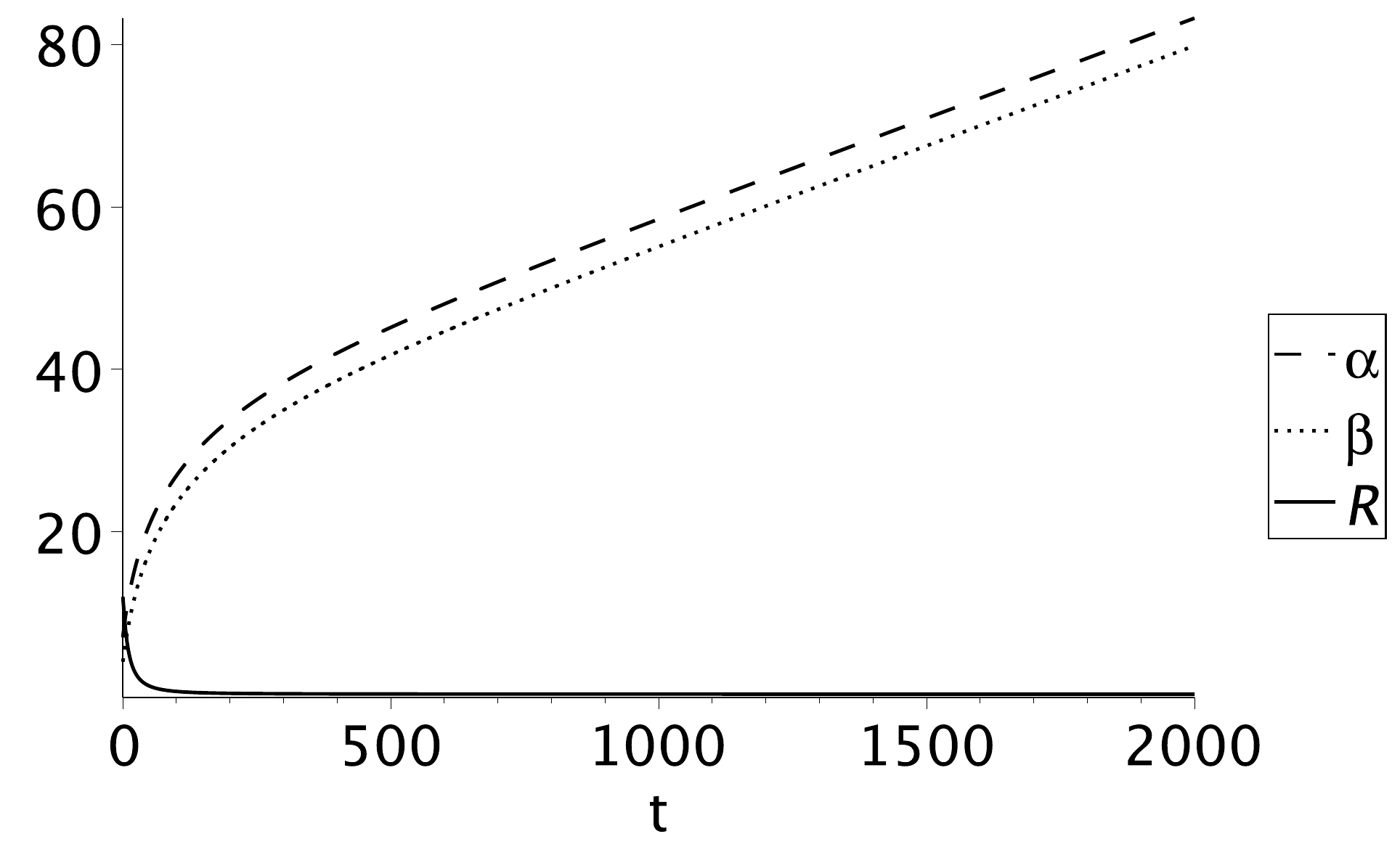} \quad \includegraphics[width=7cm]{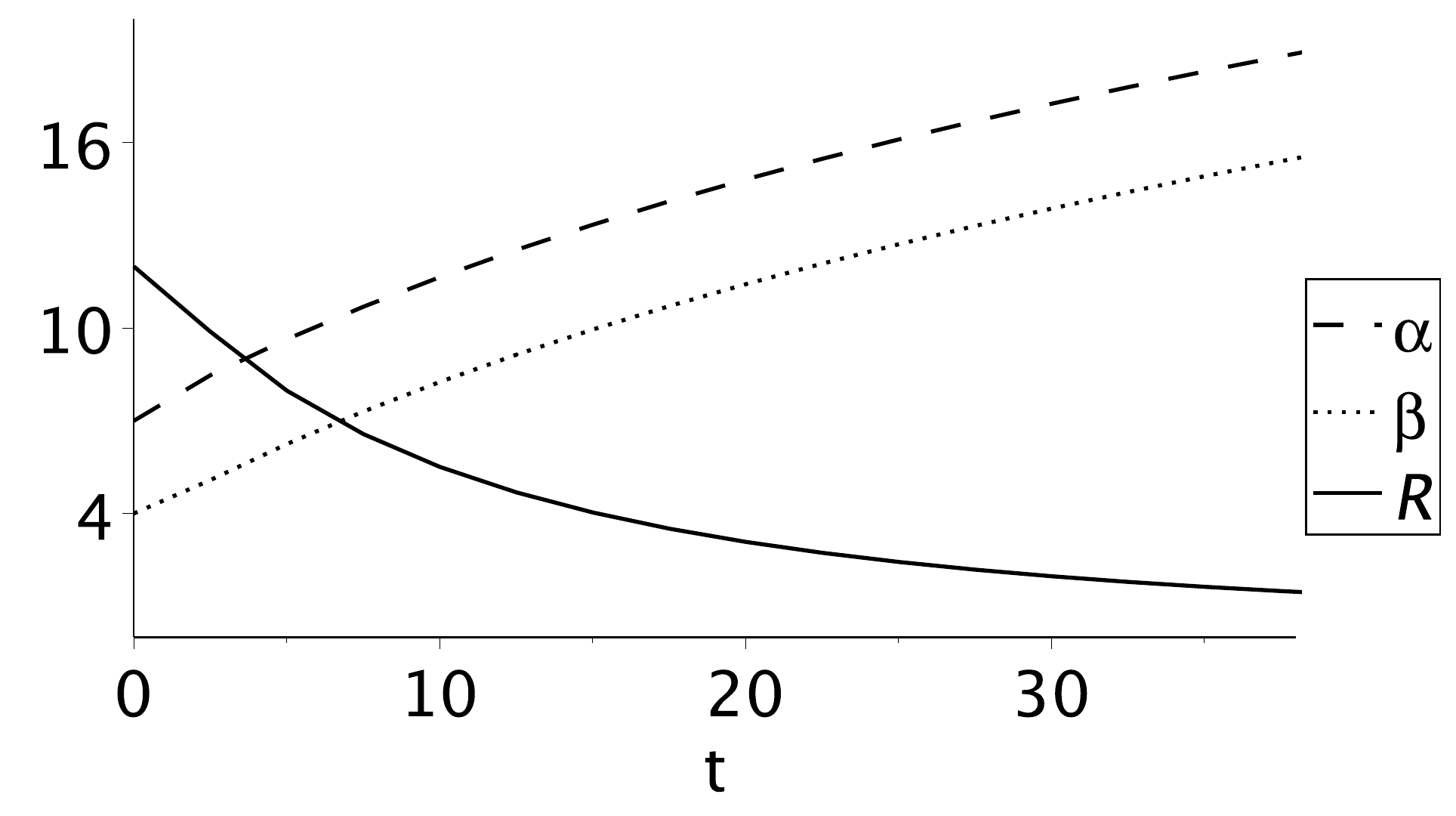}
\caption{Numerical solution to the system of Equations \eqref{xx2}--\eqref{R} for initial conditions \mbox{$ \alpha(0)=7, \ \dot{\alpha}(0)=1, \ \beta(0)=4, \ \dot{\beta}(0)=0, \ \dot{R}(0)=0$}. The~initial condition $R(0)\simeq 11.999$ is found from Equation~\eqref{eq002}.   The~Lagrangian parameters are  $ a = 200, c = -0.001, k = 500$.}
 \label{fig1}
\end{figure}

 An example of numerical solution to system \eqref{tt2}--\eqref{thth2} is represented in Figure~\ref{fig1}.

\begin{figure}[h!]
\centering
\includegraphics[width=7cm]{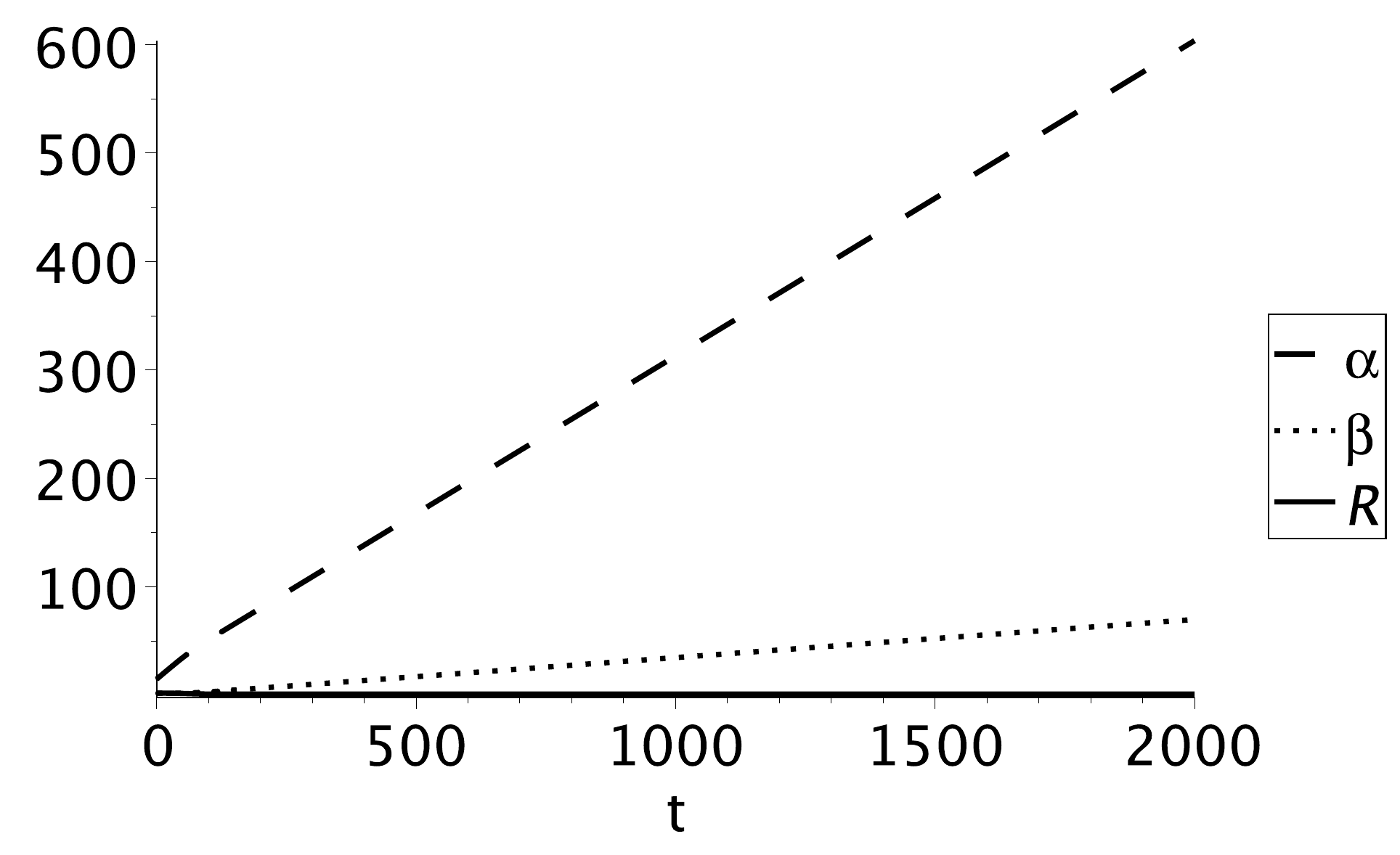} \quad \includegraphics[width=7cm]{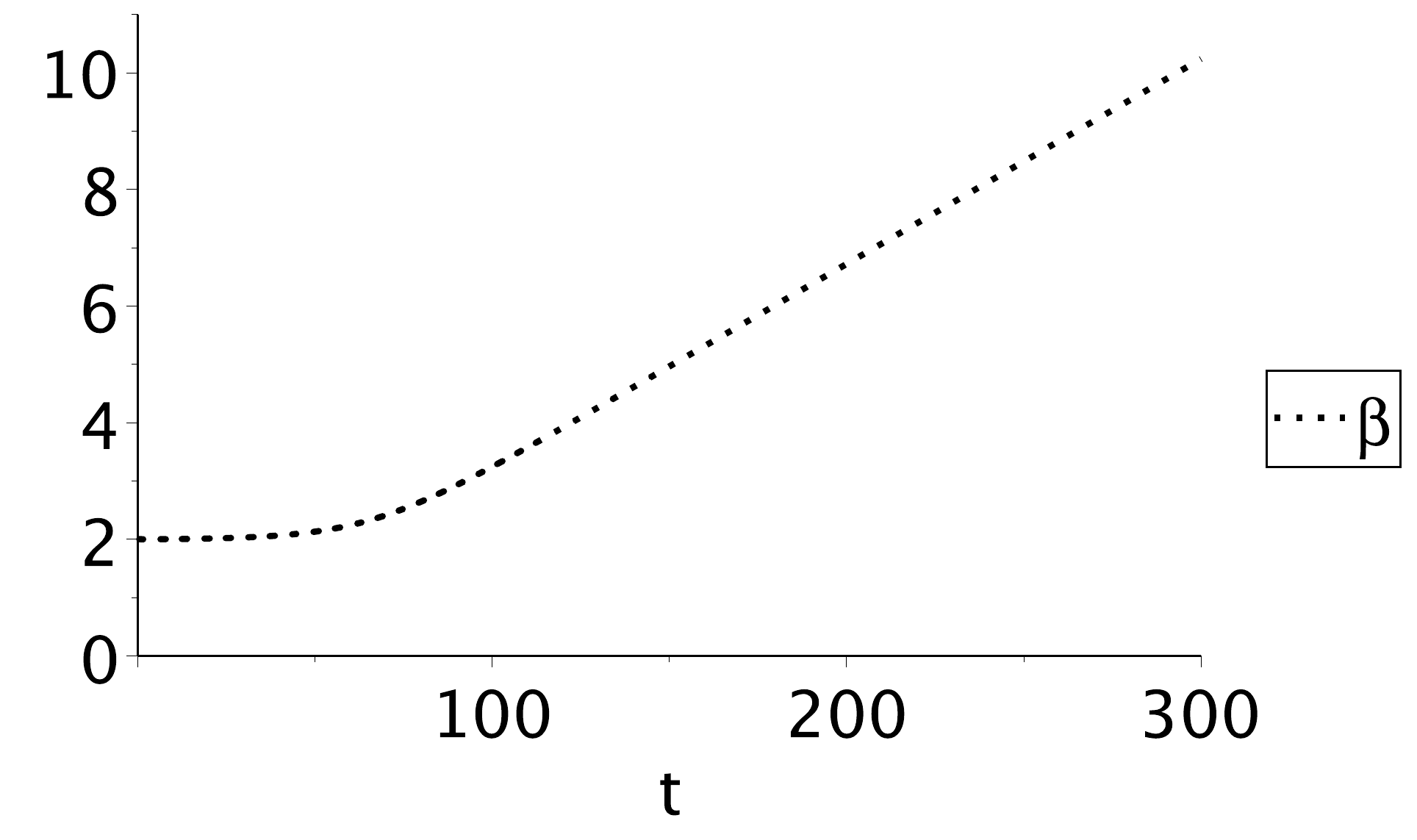}
\caption{Numerical solution to the system of Equations \eqref{xx2}--\eqref{R} for initial conditions \mbox{$ \alpha(0)=15, \ \beta(0)=2, \ \dot{\alpha}(0)\simeq 0.404667, \ \dot{\beta}(0)=0, \ \dot{R}(0)=0$}. $R(0)\simeq 2.09126$ is found from Equation~\eqref{eq002}.  The~Lagrangian parameters are $ a = -2.77, c = -0.49, k = -2.98$.}
\label{fig2_}
\end{figure}

\begin{figure}[h!]
\begin{center}
\includegraphics[width=8cm]{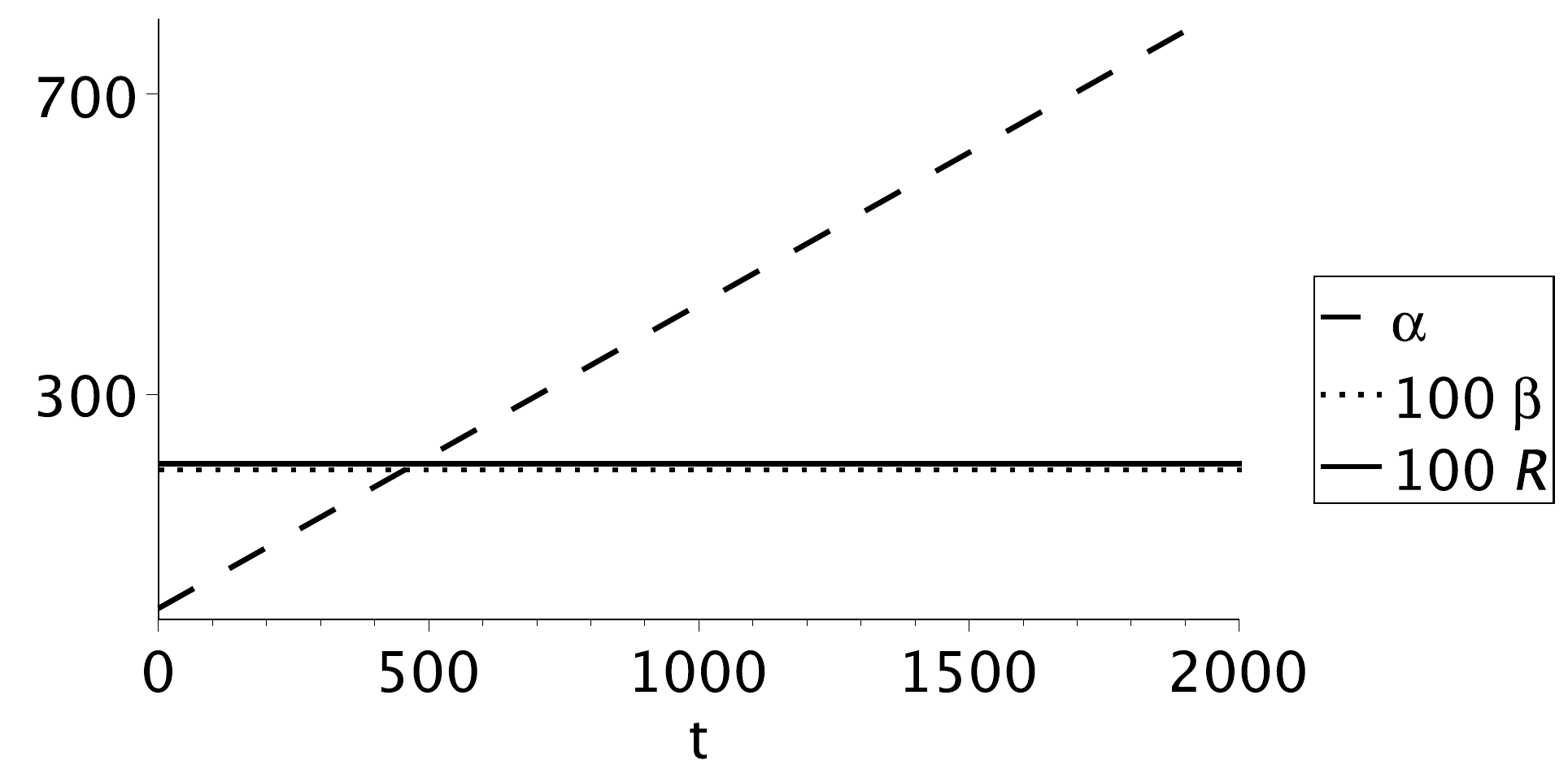}
\caption{Numerical solution to the system of Equations \eqref{xx2}--\eqref{R} for initial conditions \mbox{$ \alpha(0)=15, \ \dot{\alpha}(0) \simeq 0.40467, \ \beta(0)=b_c \simeq 1.99303$,  \ $\dot{\beta}(0)=0, \ \dot{R}(0)=0$. $R(0)\simeq 2.0765$} is found from Equation \eqref{eq002}. For~the found numerical solution, $a = -2.77, c = -0.49, k = -2.98$.}
\label{oneexpon_}
\end{center}
\end{figure}

The choice of another set of physical parameters can change the picture. Sub-spaces grow at different rates as is represented in Figures~\ref{fig2_} and~\ref{oneexpon_}.

\unskip

\unskip

\section{Approximate~Method}

The gravity with higher derivatives provides us with new abilities in understanding the role of metric in the observed world. The~price is that the analysis becomes much more complicated. The~attraction of extra dimensions aggravates the situation. Fortunately, a~small parameter arises naturally, provided that the extra space is small and compact.
Let us accept for estimation that the average Ricci scalar $R_n$ of the extra space relates to its size $r_n$ as $R_n\sim 1/r_n^2$. This means that $R_n$ is in many orders of magnitude larger than the 4-dim Ricci~scalar.

As a result, small parameter
\begin{equation}\label{ll}
\epsilon \equiv R_4/R_n \ll 1
\end{equation}
appears that strongly facilitate an analysis. Below,~we follow the method developed in~\cite{Bronnikov:2005iz}

\subsection{Basic~Idea}

Consider the gravity with higher derivatives  \eqref{act1}.
The metric is assumed to be the direct product $M_4\times V_n$ of the 4-dim space $M_4$ and $n$-dim compact space $V_n$
\begin{equation}\label{metric}
ds^2 =g_{6,AB}dz^A dz^B = g_{4,\mu\nu}(x)dx^{\mu}dx^{\mu} + g_{n,ab}(y)dy^a dy^b.
\end{equation}

Here, $g_{4,\mu\nu}$ is a metric of the manifolds $M_4$ and $g_{n,ab}(y)$ is a metric of the manifolds $V_n$.  $x$ and $y$ are the coordinates of the subspaces $M_4$ and $V_n$, respectively. We will refer to 4-dim space $M_4$ and $n$-dim compact space $V_n$ as the main space-time and an extra space, respectively. The~metric has the signature (+ - - - ...), the~Greek indexes $\mu, \nu =0,1,2,3$ refer to 4-dimensional coordinates. Latin indexes run over $a,b = 4, 5,...$.

According to \eqref{metric}, the~Ricci scalar represents a simple sum of the Ricci scalar of the main space and the Ricci scalar of extra space
\begin{equation}
R=R_4 + R_n .
\end{equation}

In this section, the~extra space is assumed to be maximally symmetric so that its Ricci scalar $R_n=const$.

Using inequality \eqref{ll} the Taylor expansion of $f(R)$ in  Equatiom \eqref{act1} gives
\begin{eqnarray}\label{act2}
&& S= \frac{m_D ^{D-2}}{2}\int d^4 x d^n y \sqrt{|g_4(x)|} \sqrt{|g_n(y)|} f(R_4 + R_n )\\
&& \simeq \frac{m_D ^{D-2}}{2}\int d^4 x d^n y\sqrt{|g_4(x)|} \sqrt{|g_n(y)|}[ R_4(x) f' (R_n) + f(R_n)]  \nonumber
\end{eqnarray}

The prime denotes derivation of functions on its argument. Thus, $f'(R)$ stands for $df/dR$ in the formula written above. Comparison of the second line in expression \eqref{act2} with the Einstein--Hilbert~action
\begin{equation}
S_{EH}=\frac{M^2_{P}}{2} \int d^4x  \sqrt{|g(x)|}(R-2\Lambda)
\end{equation}
gives the expression
\begin{equation}\label{MPl}
M^2 _{P}=m_{D} ^{D-2}v_n f' (R_n)
\end{equation}
for the Planck mass. Here, $v_n$ is the volume of the extra space. The~term
\begin{equation}\label{Lambda0}
\Lambda \equiv -\frac{m_D ^{D-2}}{2M^2 _{Pl}}v_n f(R_n)
\end{equation}
represents the cosmological $\Lambda$ term. Both the Planck mass and the $\Lambda$ term depend on a function $f(R)$. Notice that, according to \eqref{MPl}, the~Planck mass could be smaller than $D$-dim Planck mass, $M_{Pl}<m_D$ for specific functions $f$ that could lead to nontrivial consequences.

\subsection{Extension of the Model: Low~Energies}

The quantum fluctuations of metric produce all possible terms that are invariant under the coordinate transformations~\cite{Donoghue:1994dn,2007ARNPS..57..329B}.
We extend our study to effective action in the form \eqref{Seff}
\begin{equation}\label{Lgen}
S_{gen}=\frac12 m_D^{D-2}\int d^D x \sqrt{g_D}[f(R)+c_1R_{AB}R^{AB}+c_2R_{ABCD}R ^{ABCD}].
\end{equation}

Another point is that the extra space metric depends on all coordinates, not only those describing extra dimensions. Therefore, the small parameter method should be expanded for this more general~case.

In this section, we use small parameter \eqref{ll} and the conformal transformations analogous to conformal transformations \eqref{gconf} to clarify the classical behavior of system acting in $D$ dimensions.

Here, we consider a $D=4+n$ - dimensional manifold $\mathbb{M}$, having the simplest geometric structure of a direct product, $\mathbb{M}=\mathbb{M}_4 \times \mathbb{M}_n$, with~the metric
\begin{equation}\label{metric5}
ds^2 = dt^2 - e^{\alpha(x)}\Big(dr^2 +r^2(d\theta^2 +\sin^2(\theta)d\varphi^2 \Big)- e^{2\beta(x)} d\Omega_n^2.
\end{equation}

It is assumed that the extra space is $n$-dimensional maximally symmetrical manifold with positive~curvature.

The size of extra dimensions is supposed to be small compared to the size of our 4-dim space so that inequality \eqref{ll} holds and we may follow the method elaborated in~\cite{Bronnikov:2005iz, 2007CQGra..24.1261B}.
According to \eqref{metric5}, the~Ricci scalar is
\begin{equation}\label{Tailor}
    	R=R_{4} + R_{n}+ P_k; \quad P_k =  +n(n+1)(\partial{\beta})^2 +2n\square{\beta} +6 n\partial_{\mu}\alpha \partial^{\mu}{\beta}.
	\end{equation}

{The} additional inequality
\begin{equation}\label{ineq00}
 P_k\ll R_{n}
	\end{equation}
means that fluctuations of the 4-dim metric coefficient $\beta(x)$ are smooth.  More specifically,
\begin{equation}\label{ep}
   |\partial_{\mu} g_{AB}| \sim \epsilon |\partial_a  g_{AB}|, \quad \epsilon \ll 1.
\end{equation}       						

Using Formulas \eqref{metric}--\eqref{ineq00}, we can perform the Taylor decomposition of the function $f(R)$ to transform the action as
\begin{eqnarray}\label{actJ}
	&&S=\frac{1}{2}v_{n}\int d^{d_0}x \sqrt{-g_0}e^{d_2\beta_n} [f'(R_{n})R_{4} + f'(R_{n})P_k + f(R_{n})+ \nonumber \\
	&& +c_1R_{AB}R^{AB}+c_2R_{ABCD}R ^{ABCD}],   \\
	&&   R_n=\phi=n(n-1)e^{-2\beta(x)},  \label{phi} \\
 	&&R_{AB}R^{AB}= \frac{(n-1)^2}{n}e^{-4\beta}+2n(n-1)e^{-2\beta}(\square{\beta}_2 + n(\partial{\beta})^2 + 4\partial_{\mu}\alpha\partial^{\mu}{\beta})+O(\epsilon^4),  \nonumber \\
	&&R_{ABCD}R ^{ABCD}= 2\frac{n-1}{n}e^{-4\beta}+4n(n-1)e^{-2\beta}(\partial \beta)^2 +O(\epsilon^4). \nonumber
	\end{eqnarray}

The action of the form \eqref{actJ} is written in the Jordan frame. We consider this frame as a ''physical'' one which gives us the Planck mass, in~particular
\begin{equation}\label{MPl_}
	M_{P}^2=v_{n}f'(\phi_m);\quad v_{n}=\frac{2\pi^{\frac{n+1}{2}}}{\Gamma(\frac{n+1}{2})}e^{n\beta_m}.
	\end{equation}
	
Here, $\phi_m$ delivers a minimum of $V_E(\phi)$ (see definition \eqref{V_} below).

It is more familiar to work in the Einstein frame. To~this end, we have to perform the conformal transformation \eqref{gconf}
\begin{equation}\label{conform}
g_{ab}\rightarrow g_{ab}^{(E)}=e^{n\beta} |f'(\phi)|g_{ab},\quad \phi\equiv R_n = n(n-1)e^{-2\beta(x)}
\end{equation}
of the metric describing the subspace $M_{4}$.
This leads to the action in the Einstein frame in form
\begin{equation}\label{Lscalar}
S_{low}=\frac12 v_n \int d^4 x \sqrt{g_4}\ \mbox{sign}(f')[R_4 + K(\phi)(\partial \phi)^2 -2V(\phi) ],
\end{equation}
\begin{eqnarray} \label{K}
&& K(\phi)=\frac{1}{4\phi^2}\biggl[6\phi^2 (f''/f')^2 - 2n\phi(f''/f')+\frac12 n(n+2)\biggr] +\frac{c_1+c_2}{f'\phi},
\end{eqnarray}
\begin{eqnarray} \label{V_}
&&V(\phi)=-\frac{\mbox{sign}(f')}{2f'^2}\biggl[\frac{|\phi|}{n(n-1)}\biggr]^{n/2}\biggl[f(\phi)+\frac{c_V}{n}\phi^2\biggr], \quad c_V=c_1 + \frac{2c_2}{(n-1)}
\end{eqnarray}
representing specific Lagrangian of the scalar-tensor gravity~\cite{Bronnikov:2005iz}. Here, $D=4+n$ and the physical meaning of the effective scalar field $\phi$ is the Ricci scalar of the extra~space.

An important remark is necessary. The~action \eqref{actJ} describes the field evolution at high energies in the Jordan frame. When the scalar field is settled in its minimum, we should express the 4-dim Planck mass according to \eqref{MPl_}.
Simultaneously, the~scalar field is evolving during inflation, and~it is worth using the Einstein frame to facilitate analysis. In~this case, we should consider
\begin{equation}\label{m4}
    m_4\equiv\sqrt{v_n}
\end{equation}
as the effective Planck mass during the inflation in the Einstein~frame.

Let us impose restrictions that follow from the observational data at low energies. In~this case, observable value of the cosmological constant is extremely small and we neglect it. We also assume that the field $\phi$ is in its stationary state. Thus,  necessary conditions are as follows:
\begin{equation}\label{VMink}
V(\phi_m)=0;\quad V'(\phi_m)=0.
\end{equation}

The inequalities
\begin{equation}\label{ineq}
\phi_m >0,\quad f'(\phi_m)>0,\quad K(\phi_m)>0.\quad V''(\phi_m)=m^2 >0
\end{equation}
are necessary to consider the field $\phi$ as the scalar field with the standard properties.
The function $f(R)$ should be specified to make specific predictions. It is chosen in the form
\begin{equation}\label{fR}
f(R)=a R^2 + b R +c.
\end{equation}

One can easily solve algebraic Equation \eqref{VMink} with respect to
\begin{equation}\label{cond1}
\phi_m =-\frac{b}{2(a+c_V/n)}
\end{equation}
and
\begin{equation}\label{cond2}
c=\frac{b^2}{4(a+c_V/n)}.
\end{equation}

Our 4-dimensional  space-time is surely not Minkowskian so that the relation \eqref{cond2} holds~approximately.

Let's consider inequalities \eqref{ineq} in more detail. The first inequality in \eqref{ineq} leads to
\begin{equation}\label{cond6}
-\frac{b }{2 \left(a +c_V/n\right)} > 0.
\end{equation}

The second inequality in \eqref{ineq} gives
\begin{equation}\label{cond3}
f'(\phi_m)>0\rightarrow \frac{b c_V/n}{ \left(a +c_V/n\right)} > 0.
\quad \mbox{or} \quad c_V < 0 
\end{equation}

The third inequality in \eqref{ineq} gives the inequality
\begin{equation}\label{cond4}
\left[ \left( 12 a^2 +4 a c_V +c_V^2 \right)n +2c_V^2 -4 c_V (c_1 +c_2) \right]>0,
\end{equation}
and the fourth equality in \eqref{ineq} leads to
\begin{equation}\label{V2}
V''(\phi_m)=-\frac{b n^2}{c_V^2} \left(\frac{a +c_V/n}{ b }\right)^3
\left[ -\frac{b}{2 (a  +c_V/n)  n (n-1)} \right]^{n/2} \equiv m^2 > 0 \quad \mbox{i.e.} \quad b >0.
\end{equation}
As the result, the~parameters of action must satisfy the conditions
\begin{equation} \label{cond23}
b>0, c_V<0, a+c_V/n<0.
\end{equation}

The presence of a small parameter permits solving two problems. The~consideration is reduced to the standard Einstein gravity + scalar field, and~four dimensions appear naturally because the integration of the extra coordinate is trivial. The~potential of the scalar field is promising from the point of view appropriate inflationary scenario. We show below how to elaborate the inflationary~model.

\subsection{Extension of the Model: Moderate~Energies}

The study above indicates that some limits on the Lagrangian parameter values can be derived in this approach's framework. Nevertheless, freedom in the choice of the parameter values remains. This freedom that is used below can be used to obtain e.g.,~appropriate inflationary models, see~\cite{Bronnikov:2009ai,Fabris:2019ecx} for~details.

First of all, we have to specify our action \eqref{Lscalar} by definition of the function $f(R)$ \eqref{fR}.
The final form of the action is
\begin{eqnarray}
\label{l-o}
S =\frac{m_4^2}{2} \int d^4x \sqrt{-g_4}\mbox{sign}(2a\phi+b)\biggr\{R + K(\phi)\phi_{;\rho}\phi^{;\rho} - 2V(\phi)\biggl\},
\end{eqnarray}
where
\begin{equation}\label{m4-1}
    m_4=\sqrt{V_n}=\sqrt{\frac{2\pi^{\frac{n+1}{2}}}{\Gamma(\frac{n+1}{2})}},\quad m_D=1,
\end{equation}
and
\begin{eqnarray}
K(\phi) & = & \frac{1}{\phi^2(2a\phi + b)^2}\biggr\{\biggr[(6 - n + \frac{n^2}{2})a^2 + 2(c_1 + c_2)a\biggr]\phi^2 \nonumber\\
& + & \biggr[\frac{n^2}{2}ab
+ (c_1 + c_2)b\biggl]\phi + \frac{n(n+ 2)}{8}b^2\biggl\},\\
V(\phi) & = & - \frac{\mbox{sign}(2 a \phi + b)}{2 (2 a \phi + b )^2} \biggr[ \frac{\phi}{n(n - 1)} \biggl]^{n/2}\biggr\{\biggr(a + \frac{c_V}{n}\biggl)\phi^2 + b\phi + c\biggl\},
\end{eqnarray}
and
\begin{eqnarray}
c_V &=& c_ 1 + 2\frac{c_2}{n - 1}.
\end{eqnarray}

\begin{figure}[h!]
\centering
		\includegraphics[width=7cm]{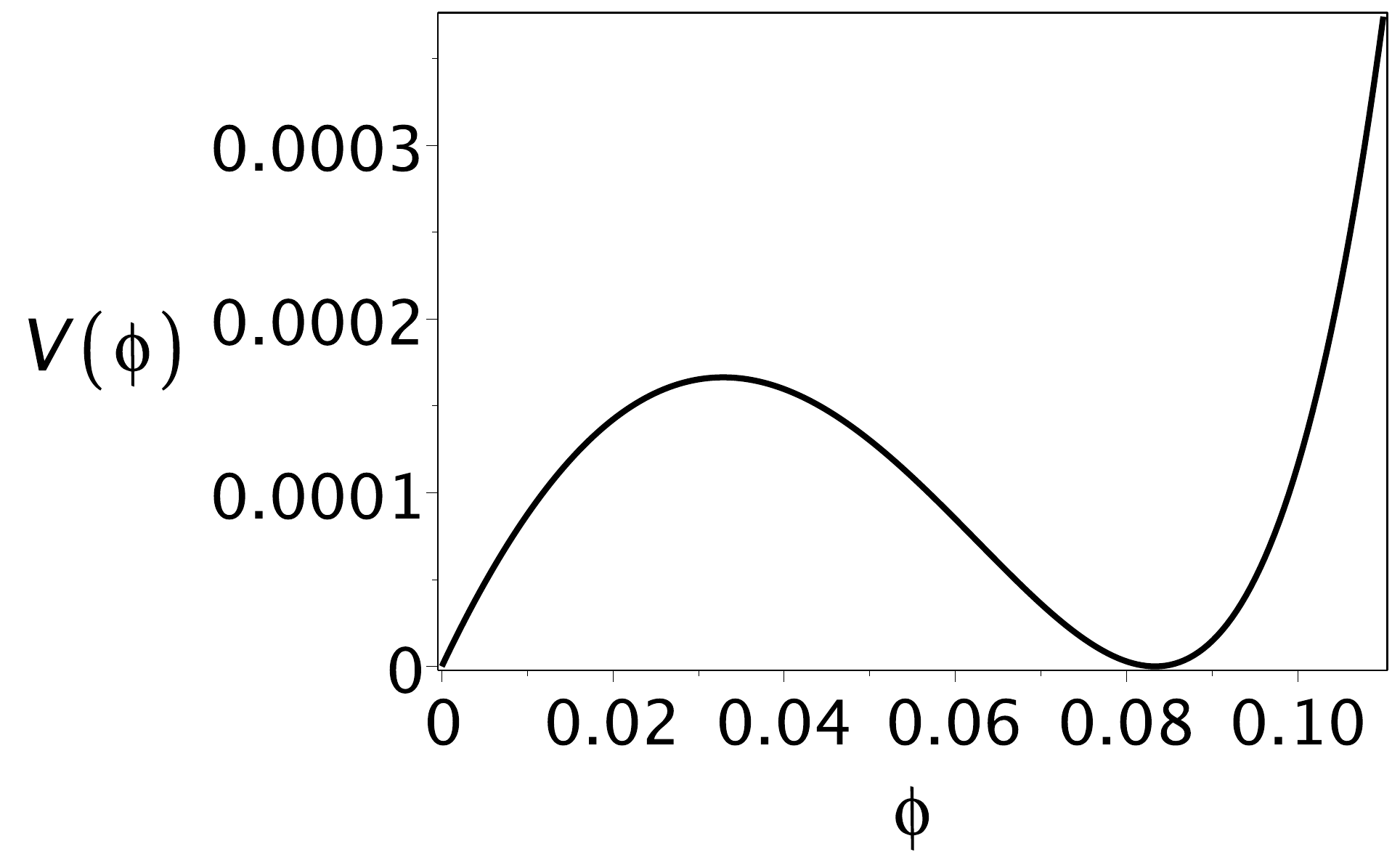} \qquad \includegraphics[width=7cm]{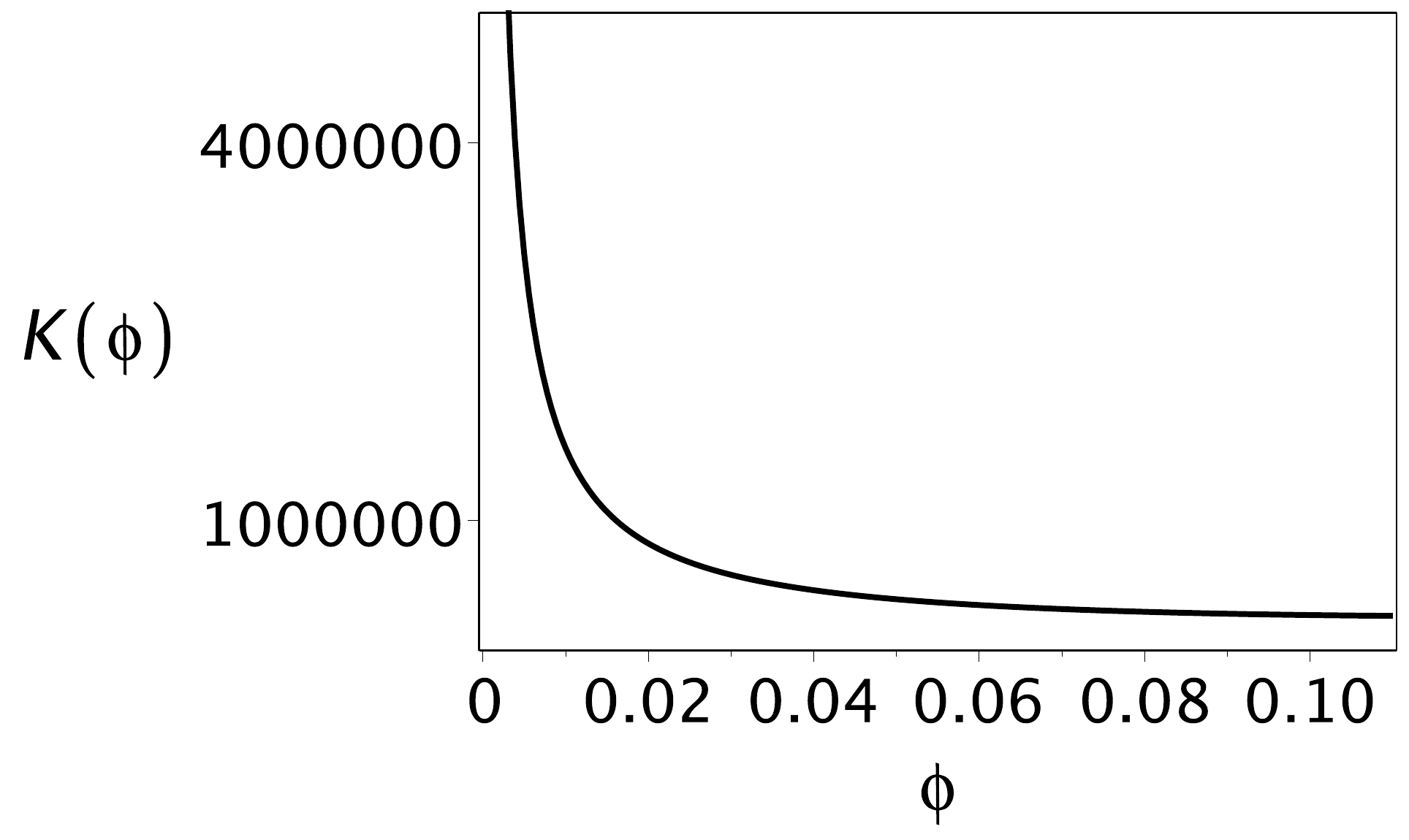}
		\caption{The form of the potential (left) and kinetic term (right) for the parameters $n = 2, b = 1, a~= -2, c_V=-8, c_K = 15000.$ The potential minimum is in the point $\phi_m  \simeq  0.083$, $m_D=1.$}\label{VK}
	\end{figure}

The kinetic factor $K(\phi)$ and the potential  $K(\phi)$ have a complex form depending on several parameters. They are represented in Figure~\ref{VK} for the parameter values
\begin{eqnarray}\label{param}
&&n = 2, b = 1, a~= -2, c_V=-8, c_K=15000,  \\
&&\left(c_V = c_ 1 + 2\frac{c_2}{n - 1}, c_K=c_1+c_2. \right) \nonumber 
\end{eqnarray}

The parameter $"c"$ can be obtained from expression \eqref{cond2}. Restrictions \eqref{ineq}--\eqref{V2} are also taken into~account.

Finally, we must check inequality \eqref{ll} (written in the Jordan frame) to be sure in the self-consistency of our approach. The~n-dim scalar Ricci is not transformed under conformal transformations \eqref{conform} so that $R_2=\phi$. Let us estimate the 4-dim part of the Ricci scalar $R_4$ keeping in mind the slow motion regime which is inherent for the inflationary period. In~this case, $R_4^{(E)}=12H^2\simeq 12\cdot 8\pi V(\phi)/3M_{Pl}^2$ in the Einstein frame. A relation to the Jordan frame may be found in \eqref{gconf}, \eqref{Rconf}, and \eqref{conform}. Finally,
\begin{equation}
    \epsilon \equiv R_4^{(J)}/R_n\sim 100f'(\phi)V(\phi)/\phi^2.
\end{equation}

Our approximation \eqref{ll} holds if the field $\phi$ is varied near the potential minimum, $\phi=\phi_m$, in the region where the potential is limited from above
\begin{equation}\label{Vlim}
    V(\phi)\ll \frac{\phi_m^2}{100 f'(\phi_m)} \sim 10^{-4}
\end{equation}
Here, $\phi_m =0.083$ in the units $m_D=1$ according to the figure above. As~we will see just below, $m_D\sim 0.1\, M_{Pl}$ so that the limitation \eqref{Vlim} is not very restricted: $V\ll 10^{-8}M_{Pl}^4$.

In the main text above, we express all quantities in units $m_D=1$ with the effective Planck mass $m_4$ defined by \eqref{m4}. Finally, when the scalar field $\phi$ has been settled in the minimum of its potential $V(\phi)$, one should restore more physical units valid for the Jordan frame, i.e.,~the Planck units using the relation \eqref{MPl_}. For~chosen parameter values \eqref{param}, the relation
\begin{equation}
M_{Pl}=\sqrt{V_{n}e^{n\beta_m}f'(\phi_m)}m_D
\end{equation}
can be used to obtain the value of the D-dimensional Planck mass.
For two-dimensional extra space ($n=2,\quad V_2 = 4\pi$),
\begin{equation}
M_{Pl}=\sqrt{V_{2}e^{2\beta_m}f'(\phi_m)}m_D = \sqrt{8\pi (2a+\frac{1}{\phi_m})}m_D \sim 10 m_D,
\end{equation}
which means that the D-dim Planck mass $m_D$ is in the order of magnitude smaller than the Planck mass in the Jordan~frame.

One can see that the presence of a small parameter strongly facilitates the analysis and could lead to promising results, even if the initial system is quite complicated. Indeed, it is hard even to guess that action \eqref{Lgen} could adequately describe the inflationary stage. On~the other side, the~energy density at the moderate energies (inflation) is not small, so that one must check the smallness of the parameter value \eqref{ll}.

\section{Method of Trial~Functions}

This method is applied if a system in question is too complicated, see, e.g.,~\cite{PhysRevD.60.067504,Boehmer2005DynamicalIO}.
The~mathematically correct way to deal with gravitational systems is as follows. One should type equations of motion for a metric of the most general form, and~solves it for desired initial and boundary conditions. The~freedom in the coordinate choice facilitates the analysis, but~the system often remains too complicated to be solved. The~problem is usually solved by choosing a limited set of metrics that is substituted into an action. Classical equations are obtained by variation of the metric belonging to the chosen set. The~well-known example is the minisuperspace model of Universe creation. The~method of trial functions is one of such a sort. Its essence is to choose an appropriate set of metrics---some functions depending on parameters. The~action appears to be dependent on the set of unknown parameters. The~classical system of equations is reduced to an algebraic system for these~parameters.

Let us illustrate it by the specific solution that non trivially connects two sub-spaces.
%
%
%
Consider a manifold $M$ with topology $T \times M_1 \times M_2 \times M_3$, where $M_1$ is one-dimensional, infinite space, and~$M_2, M_3$ are two-dimensional spheres. We study the pure gravitational field action in the form (\ref{Lgen}) with the metric
\begin{equation}
\label{eq:g_fun}
ds^2 =  A(u) dt^2 - A(u)^{-1} du^2 - e^{2\beta_1 (u)} d\Omega_1^2 - e^{2\beta_2 (u)} d \Omega_2^2.
\end{equation}

Here, $A(u)$, $\beta_1(u)$, and $\beta_2(u)$ are members of the limited set of metrics depending on the Schwarzschild radial coordinate $u$, $-\infty < u < \infty$. The~action contains invariants depending on these functions. For~example, the~Ricci scalar has the form (see~\cite{2016PhLB..759..622R,Lyakhova:2018zsr} for details)
$$
R = 2 e^{-2\beta_1}+2e^{-2\beta_2}-A^{\prime \prime} - 4 A^{\prime} \left( \beta_1^{\prime} + \beta_2^{\prime}\right)$$
\begin{equation}
\label{eq:R_sc}
- 2 A \left(3\beta_1^{\prime 2}+3\beta_2^{\prime 2} + 4 \beta_1^{\prime} \beta_2^{\prime} + 2 \beta_1^{\prime \prime} + 2 \beta_2^{\prime \prime} \right),
\end{equation}
with prime denoting differentiation with respect to $u$.

Let us study the metric that represents the transition between the domain with (large $ M_2 $/small $ M_1 $) subspaces and the domain with subspaces (large $ M_1 $/small $ M_2 $)  \cite{2016PhLB..759..622R} with the Minkowski metric at the asymptotes:
\begin{equation}\label{A}
A(u \to \pm \infty) \rightarrow 1,
\end{equation}
and
\begin{eqnarray}\label{b1}
&&\beta_1(u \to + \infty) \rightarrow \ln u, \qquad \beta_1(u \to -\infty) \rightarrow \ln r_0,\\
&&\beta_2(u \to - \infty) \rightarrow \ln u, \qquad \beta_2(u \to +\infty) \rightarrow \ln r_0. \nonumber
\end{eqnarray}

Here, $r_0=e^{\beta_c}$ is the radius of extra space at $u\rightarrow \pm \infty$. The~value of $r_0 = 1/\sqrt{-c}$ is connected with the physical parameter $c$ according to Formulas (\ref{phi})--(\ref{cond2}).

For the numerical simulations, we  use the Ritz method which means that the functions $A(u), \beta_1(u), \beta_2(u)$ in (\ref{eq:g_fun}) are approximated by trial functions that are chosen in the form
\begin{equation}
\label{eq:trial}
A(u) = 1 - \frac{\xi_2}{\sqrt{\xi_1^2 + u^2}},
\end{equation}
\begin{displaymath}
e^{\beta_1(u)} = r_0+\frac{1}{2} \left(u+\sqrt[4]{\xi_3^4+u^4}\right),
\end{displaymath}
\begin{displaymath}
e^{\beta_2(u)} = r_0-\frac{1}{2} \left(u-\sqrt[4]{\xi_3^4+u^4}\right),
\end{displaymath}
keeping in mind conditions \eqref{A}, \eqref{b1}. According to~\cite{Lyakhova:2018zsr}, the~parameters $\xi_1, \xi_2, \xi_3$ should satisfy~equation
\begin{equation}
\label{eq:cond22}
\left(\frac{2\xi_2}{c}+\xi_1^2\xi_2+3\sqrt{-c}\xi_3^4\right) (1-4c^2)+3c_1(-c)^{3/2} \xi_3^4 = 0.
\end{equation}

In what follows, we consider $\xi_1,\xi_3$ as independent values, and~$\xi_2$ as their function. Parameters $\xi_1, \xi_2, \xi_3$ are to be defined by the action minimization.


Now, the action $S(\xi_1,\xi_3)$ appears to be a function of two variables after the substitution of trial functions \eqref{eq:trial} into action \eqref{Lgen}. Classical solutions should satisfy equations
\begin{equation}
\frac{\delta S}{\delta g_{ab}}=0
\end{equation}
or, {according to the Ritz method,} \cite{doi:10.1137/100804036}
\begin{equation}\label{appsol}
\frac{\partial S}{\delta \xi_1}=0,\quad \frac{\partial S}{\delta \xi_3}=0.
\end{equation}

One of the ways to solve this system is to find a minimum of the auxiliary function
\begin{equation}
\label{eq:FUN}
\Omega(\xi_1, \xi_2(\xi_1,\xi_3), \xi_3) = \left( \frac{\partial S}{\partial \xi_1} \right)^2 + \left( \frac{\partial S}{\partial \xi_3} \right)^2.
\end{equation}

\begin{figure}[h!]
\centering
\includegraphics[scale=0.4]{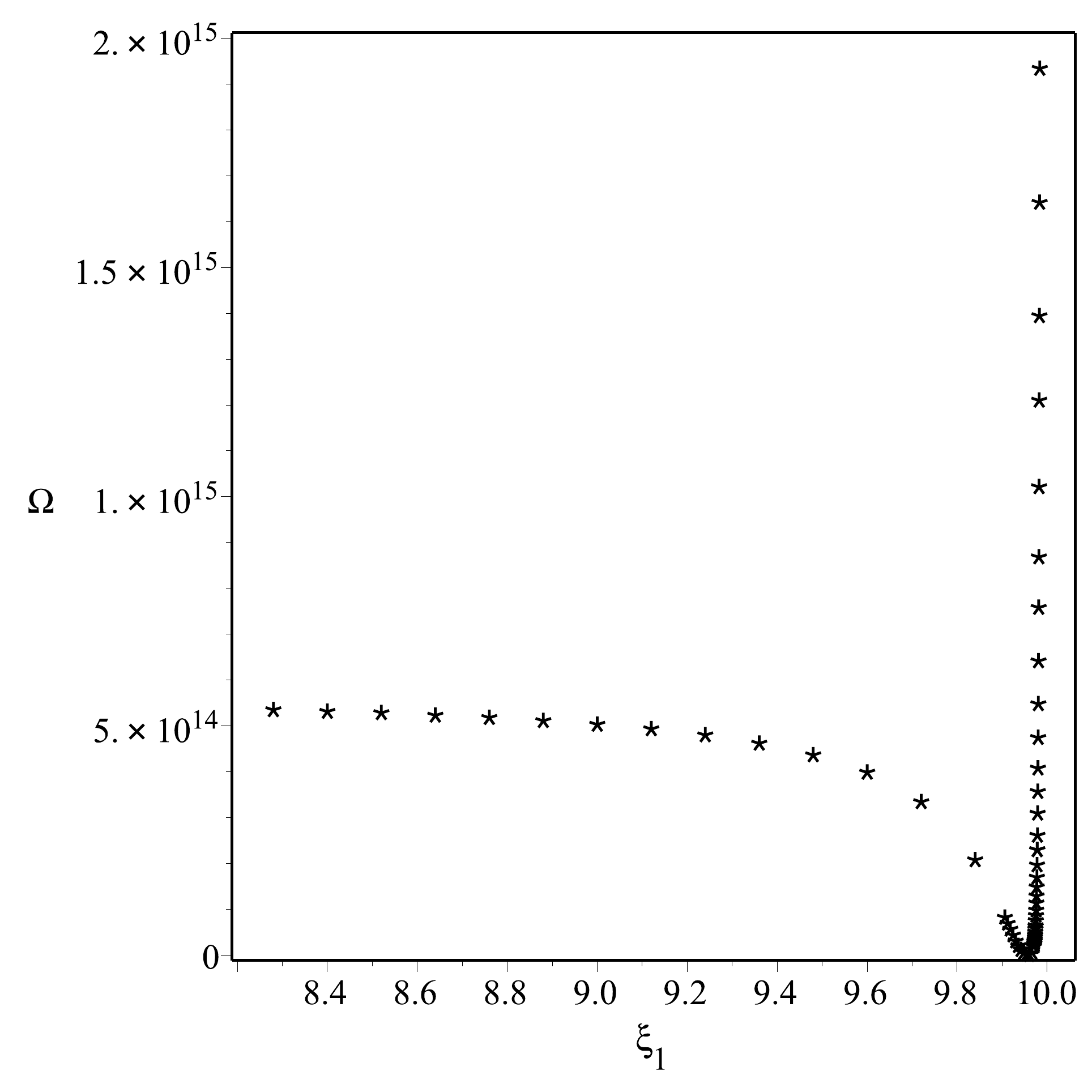}
\caption{Auxiliary function $\Omega$ vs. the minimization parameter $\xi_1$ for $\xi_2^*=9.57,\xi_3^*=1.85$ and $f(R)$-parameters $a = -2, b = 1, c = -0.02$. The~minimum of $\Omega$ corresponds to $\xi_1^* = 9.96$. $\Omega (\xi_1^{*} = 9.96, \xi_3^* = 1.85)= 5.3 \times 10^{11}$.}
\label{fig:FUN_xi1}
\end{figure}

\begin{figure}[h!]
\centering
\includegraphics[scale=0.3]{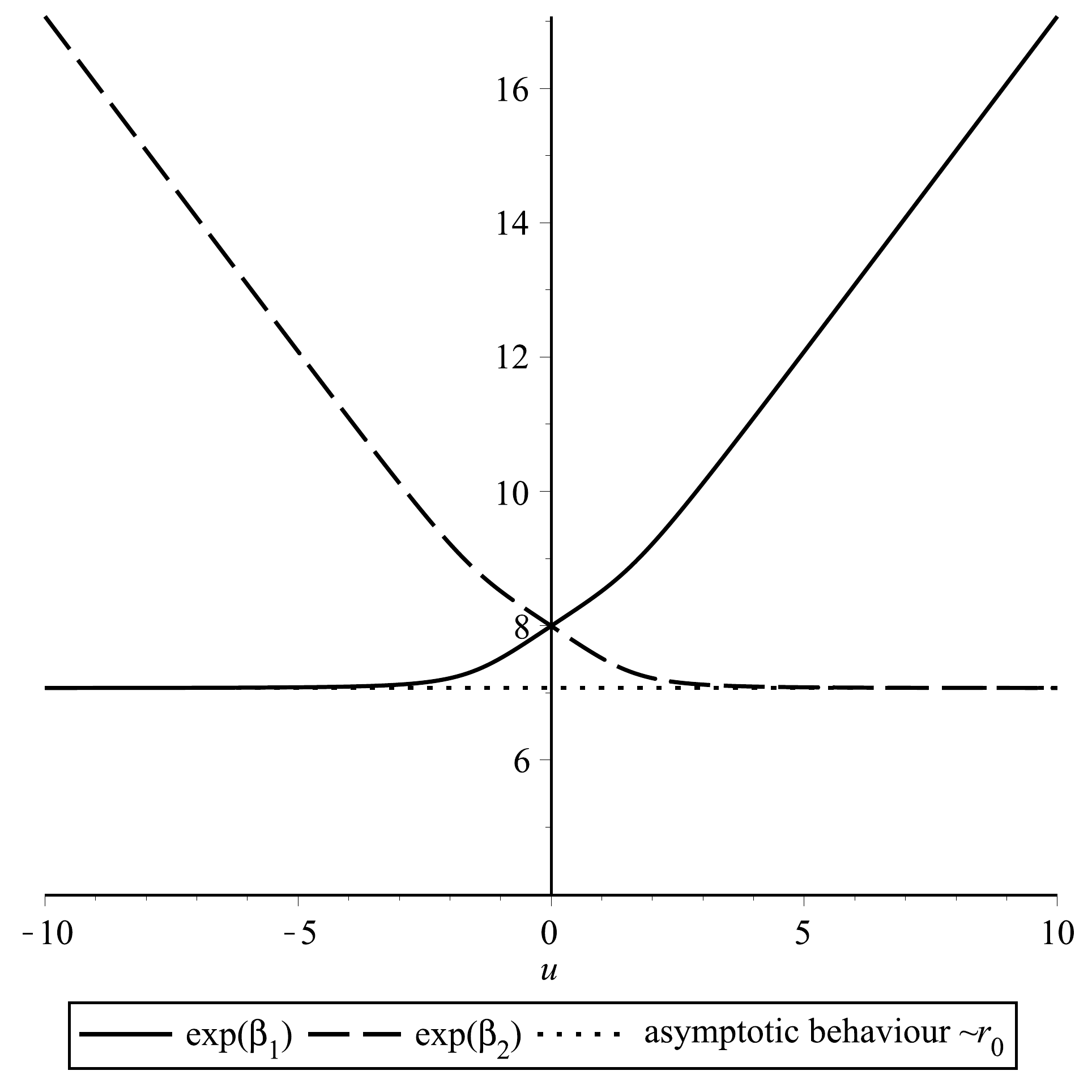}
\caption{Radii of two-dimensional subspaces vs. the Schwarzschild radial coordinate $u$ for parameter values  $a = -2, b = 1, c = -0.02$, $\xi_3 = 1.85$.}
\label{fig:FUN_8}
\end{figure}

The minimization procedure for $\Omega(\xi_1,\xi_2(\xi_1,\xi_3),\xi_3)$ gives the minimum at $\xi_1^{*} = 9.96,\xi_2^*=9.57,\xi_3^*=1.85$, see Figure~\ref{fig:FUN_xi1}. The minimum is very profound so that the trial functions, see Figure~\ref{fig:FUN_8}, are chosen~properly.

\unskip


For an external observer, the~funnel looks like a microscopic object with the mass of the Planck scale's order. 

The method of trial functions (the Ritz method) is a useful tool to study complicated systems provided that the boundary conditions are~known.

\section{Conclusions}

In this review, we discussed the working methods that help to cope with the mathematical difficulties of gravity with higher derivatives. All methods are supplied by examples that help to reveal the positive and negative aspects of each~approach.

The method based on the conformal transformation strongly facilitates analysis in some cases, but~it is mostly limited by $f(R)$ models. In addition, caution in the application of this method is necessary if the quantum effects are~essential.

The direct solution of the system of differential equations followed from an action minimization is a complicated problem. In addition,  the~progress is possible if these equations are of the second order in time derivatives. The~latter is not obligatorily true for models containing invariants other than the Ricci~scalar.

If the system contains a small parameter, this usually helps to obtain results. The~ratio of the Ricci scalars of extra dimensions and those measured in the present Universe can be used in many models containing the extra space. It often helps to reduce a primary D-dim action with higher derivatives to the 4-dimensional action with a scalar field. The~smallness of such ratio \eqref{ll} should be controlled at high energies~intentionally.

The trial functions method is quite universal, but~the accuracy of results is not evident, and~efforts should be applied to clarify the~question.

\section{Acknowledgments}

This research was funded by the Ministry of Science and Higher Education of the Russian Federation, Project ``Fundamental properties of elementary particles and cosmology'' N 0723-2020-0041.
The work of A.P.  and S.R. is performed according to the Russian Government Program of Competitive Growth of Kazan Federal University. The~work of A.P. was partly funded by the Russian Foundation for Basic Research Grant No. 19-02-00496.
The work of A.P was also funded by the development program of the Regional Scientific and Educational Mathematical Center of the Volga Federal District, agreement N 075-02-2020.


\end{document}